\documentclass[pra,aps,onecolumn,twoside,superscriptaddress]{revtex4}

\usepackage{amsmath,amsfonts,amssymb,caption,color,epsfig,graphics,graphicx,hyperref,latexsym,mathrsfs,revsymb,theorem,url,verbatim,epstopdf,mathtools,enumerate,graphicx, float, extarrows, tcolorbox}

\usepackage{listings}
\usepackage{color}

\definecolor{dkgreen}{rgb}{0,0.6,0}
\definecolor{gray}{rgb}{0.5,0.5,0.5}
\definecolor{mauve}{rgb}{0.58,0,0.82}

\lstdefinelanguage{JavaScript}{
  keywords={typeof, new, true, false, catch, function, return, null, catch, switch, var, if, in, while, do, else, case, break},
  keywordstyle=\color{blue}\bfseries,
  ndkeywords={class, export, boolean, throw, implements, import, this},
  ndkeywordstyle=\color{darkgray}\bfseries,
  identifierstyle=\color{black},
  sensitive=false,
  comment=[l]{//},
  morecomment=[s]{/*}{*/},
  commentstyle=\color{purple}\ttfamily,
  stringstyle=\color{red}\ttfamily,
  morestring=[b]',
  morestring=[b]"
}

\lstdefinelanguage{HTML5}{
        language=html,
        sensitive=true, 
        alsoletter={<>=-},
        otherkeywords={
        <html>, <head>, <title>, </title>, <meta, />, </head>, <body>,
        <canvas, \/canvas>, <script>, </script>, </body>, </html>, <!, html>, <style>, </style>, ><
        },  
        ndkeywords={
        =,
        charset=, id=, width=, height=,
        border:, transform:, -moz-transform:, transition-duration:, transition-property:, transition-timing-function:
        },  
        morecomment=[s]{<!--}{-->},
        tag=[s]
}

\newtheorem{definition}{Definition}
\newtheorem{proposition}{Proposition}
\newtheorem{lemma}{Lemma}

\newtheorem{theorem}{Theorem}
\newtheorem{corollary}{Corollary}
\newtheorem{conjecture}{Conjecture}

\newtheorem{remark}{Remark}
\newtheorem{example}{Example}
\newtheorem{question}{Question}
\newtheorem{memo}{Memo}
\newtheorem{scheme}{Scheme}

\def\squareforqed{\hbox{\rlap{$\sqcap$}$\sqcup$}}
\def\qed{\ifmmode\squareforqed\else{\unskip\nobreak\hfil
\penalty50\hskip1em\null\nobreak\hfil\squareforqed
\parfillskip=0pt\finalhyphendemerits=0\endgraf}\fi}
\def\endenv{\ifmmode\;\else{\unskip\nobreak\hfil
\penalty50\hskip1em\null\nobreak\hfil\;
\parfillskip=0pt\finalhyphendemerits=0\endgraf}\fi}
\newenvironment{proof}{\noindent \textbf{{Proof.~} }}{\qed}
\def\Dbar{\leavevmode\lower.6ex\hbox to 0pt
{\hskip-.23ex\accent"16\hss}D}
\makeatletter
\def\url@leostyle{%
  \@ifundefined{selectfont}{\def\UrlFont{\sf}}{\def\UrlFont{\small\ttfamily}}}
\makeatother
\def\bcj{\begin{conjecture}}
\def\ecj{\end{conjecture}}
\def\bcr{\begin{corollary}}
\def\ecr{\end{corollary}}
\def\bd{\begin{definition}}
\def\ed{\end{definition}}
\def\bea{\begin{eqnarray}}
\def\eea{\end{eqnarray}}
\def\bem{\begin{enumerate}}
\def\eem{\end{enumerate}}
\def\bex{\begin{example}}
\def\eex{\end{example}}
\def\bim{\begin{itemize}}
\def\eim{\end{itemize}}
\def\bl{\begin{lemma}}
\def\el{\end{lemma}}
\def\bma{\begin{bmatrix}}
\def\ema{\end{bmatrix}}
\def\bpf{\begin{proof}}
\def\epf{\end{proof}}
\def\bpp{\begin{proposition}}
\def\epp{\end{proposition}}
\def\bqu{\begin{question}}
\def\equ{\end{question}}
\def\br{\begin{remark}}
\def\er{\end{remark}}
\def\bt{\begin{theorem}}
\def\et{\end{theorem}}
\def\bmm{\begin{memo}}
\def\emm{\end{memo}}

\def\btb{\begin{tabular}}
\def\etb{\end{tabular}}

\newcommand{\nc}{\newcommand}

 \nc{\bbA}{\mathbb{A}} \nc{\bbB}{\mathbb{B}} \nc{\bbC}{\mathbb{C}}
 \nc{\bbD}{\mathbb{D}} \nc{\bbE}{\mathbb{E}} \nc{\bbF}{\mathbb{F}}
 \nc{\bbG}{\mathbb{G}} \nc{\bbH}{\mathbb{H}} \nc{\bbI}{\mathbb{I}}
 \nc{\bbJ}{\mathbb{J}} \nc{\bbK}{\mathbb{K}} \nc{\bbL}{\mathbb{L}}
 \nc{\bbM}{\mathbb{M}} \nc{\bbN}{\mathbb{N}} \nc{\bbO}{\mathbb{O}}
 \nc{\bbP}{\mathbb{P}} \nc{\bbQ}{\mathbb{Q}} \nc{\bbR}{\mathbb{R}}
 \nc{\bbS}{\mathbb{S}} \nc{\bbT}{\mathbb{T}} \nc{\bbU}{\mathbb{U}}
 \nc{\bbV}{\mathbb{V}} \nc{\bbW}{\mathbb{W}} \nc{\bbX}{\mathbb{X}}
 \nc{\bbZ}{\mathbb{Z}}

 \nc{\bA}{{\bf A}} \nc{\bB}{{\bf B}} \nc{\bC}{{\bf C}}
 \nc{\bD}{{\bf D}} \nc{\bE}{{\bf E}} \nc{\bF}{{\bf F}}
 \nc{\bG}{{\bf G}} \nc{\bH}{{\bf H}} \nc{\bI}{{\bf I}}
 \nc{\bJ}{{\bf J}} \nc{\bK}{{\bf K}} \nc{\bL}{{\bf L}}
 \nc{\bM}{{\bf M}} \nc{\bN}{{\bf N}} \nc{\bO}{{\bf O}}
 \nc{\bP}{{\bf P}} \nc{\bQ}{{\bf Q}} \nc{\bR}{{\bf R}}
 \nc{\bS}{{\bf S}} \nc{\bT}{{\bf T}} \nc{\bU}{{\bf U}}
 \nc{\bV}{{\bf V}} \nc{\bW}{{\bf W}} \nc{\bX}{{\bf X}}
 \nc{\bZ}{{\bf Z}}

\nc{\cA}{{\cal A}} \nc{\cB}{{\cal B}} \nc{\cC}{{\cal C}}
\nc{\cD}{{\cal D}} \nc{\cE}{{\cal E}} \nc{\cF}{{\cal F}}
\nc{\cG}{{\cal G}} \nc{\cH}{{\cal H}} \nc{\cI}{{\cal I}}
\nc{\cJ}{{\cal J}} \nc{\cK}{{\cal K}} \nc{\cL}{{\cal L}}
\nc{\cM}{{\cal M}} \nc{\cN}{{\cal N}} \nc{\cO}{{\cal O}}
\nc{\cP}{{\cal P}} \nc{\cQ}{{\cal Q}} \nc{\cR}{{\cal R}}
\nc{\cS}{{\cal S}} \nc{\cT}{{\cal T}} \nc{\cU}{{\cal U}}
\nc{\cV}{{\cal V}} \nc{\cW}{{\cal W}} \nc{\cX}{{\cal X}}
\nc{\cZ}{{\cal Z}}

\nc{\hA}{{\hat{A}}} \nc{\hB}{{\hat{B}}} \nc{\hC}{{\hat{C}}}
\nc{\hD}{{\hat{D}}} \nc{\hE}{{\hat{E}}} \nc{\hF}{{\hat{F}}}
\nc{\hG}{{\hat{G}}} \nc{\hH}{{\hat{H}}} \nc{\hI}{{\hat{I}}}
\nc{\hJ}{{\hat{J}}} \nc{\hK}{{\hat{K}}} \nc{\hL}{{\hat{L}}}
\nc{\hM}{{\hat{M}}} \nc{\hN}{{\hat{N}}} \nc{\hO}{{\hat{O}}}
\nc{\hP}{{\hat{P}}} \nc{\hR}{{\hat{R}}} \nc{\hS}{{\hat{S}}}
\nc{\hT}{{\hat{T}}} \nc{\hU}{{\hat{U}}} \nc{\hV}{{\hat{V}}}
\nc{\hW}{{\hat{W}}} \nc{\hX}{{\hat{X}}} \nc{\hZ}{{\hat{Z}}}

\nc{\hn}{{\hat{n}}}

\def\dim{\mathop{\rm Dim}}

\def\max{\mathop{\rm max}}
\def\min{\mathop{\rm min}}

\def\sn{\mathop{\rm SN}}

\def\ox{\otimes}

\newcommand{\bra}[1]{\langle#1|}
\newcommand{\ket}[1]{|#1\rangle}
\newcommand{\proj}[1]{| #1\rangle\!\langle #1 |}

\def\Dbar{\leavevmode\lower.6ex\hbox to 0pt
{\hskip-.23ex\accent"16\hss}D}

\begin{document}
\title{Investigation of the PPT Squared Conjecture for High Dimensions}

\date{\today}


\author{Ryan Jin}
\email{ryan03px2021@saschina.org}
\affiliation{Shanghai American School Puxi}
\affiliation{}

\author{Yu Yang}
\email{yy19900320@icloud.com}

\begin{abstract}
We present the positive-partial-transpose squared conjecture introduced by M. Christandl at Banff International Research Station Workshop: Operator Structures in Quantum Information Theory (Banff International Research Station, Alberta, 2012) \cite{report2012}. We investigate the conjecture in higher dimensions and offer two novel approaches (decomposition and composition of quantum channels) and correspondingly, several schemes for finding counterexamples to this conjecture. One of the schemes involving the composition of PPT quantum channels in unsolved dimensions yields a potential counterexample.\\

\end{abstract}

\maketitle

\tableofcontents
\newpage
\section{Introduction}
\label{sec:int}
In a world that is increasingly characterized by the plethora of technology and data, privacy and security become an increasingly relevant issue. Quantum information theory and quantum teleportation leverage the peculiar properties of quantum entanglement to establish secure channels of communication, transmitting quantum information of a quantum system to another location. To extend the range of communication, a quantum repeater between the sender and the receiver is often used. Consider the following figure

\begin{figure}[H]
  \includegraphics[width=\linewidth]{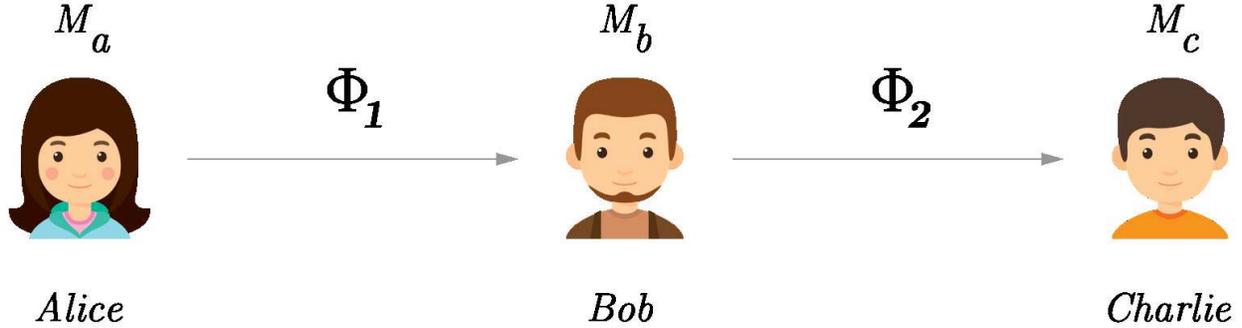}
  \caption{Quantum Communication Amongst Alice, Bob, and Charlie}
\end{figure}

$\phi_1$ is a quantum channel between Alice and Bob, and $\phi_2$ is a quantum channel between Bob and Charlie. Suppose Alice has a particle that is entangled with a particle that belongs to Bob, and Bob teleports the quantum information of his particle to Charlie. Although Alice has never interacted with Charlie, their particles are now entangled through the composite channel $\phi_1\circ\phi_2$. The process of getting $\phi_1\circ\phi_2$ is called entanglement swapping. It is a critical element of quantum repeaters as it establishes a secret key over a long distance by maintaining quantum entanglement over short distances.

While noises affect the communication of classical channels (take telephone for an example), it can also decrease the extent to which a quantum channel can maintain quantum entanglement. Moreover, the composition of multiple quantum channels also increases the likelihood of noises that break the quantum entanglement. As a result, it is natural to inquire when are composite quantum channels entanglement-breaking. Referencing the figure above, when can Alice communicate with Charlie through the composite channel $\phi_1(\phi_2)$, and when can she not?
\subsection{A Brief Overview of Quantum States and Linear Maps}
In a nutshell, a \textbf{quantum state} is a vector that encodes the state and contains the information of a system. However, due to the Heisenberg uncertainty principle, only some of the information could be extracted at a time (e.g. exact measurement of the position and momentum cannot be known simultaneously). In addition, a quantum system can be in a mixture of states simultaneously; that is known as ``quantum superposition", and such a mixture of quantum states is called a \textbf{mixed state}. A quantum state that can be expressed using a single vector (i.e. cannot be expressed as a mixture of states) is called a \textbf{pure state}. Both types of quantum states can be expressed by a \textbf{density matrix}. The formulations of quantum systems, quantum states, and density matrices are mathematically characterized below:

\bd[Quantum System and Quantum State]
A finite-dimensional \textbf{quantum system} with n states is represented by an n-dimensional complex space $\bbC^n$. On the other hand, a quantum state $\ket{\phi}$ is a vector in $\bbC^n$.

\ed

Let $\mathcal{H}_{A}$ and $\mathcal{H}_{B}$ denote two finite-dimensional quantum systems, where $\dim(\mathcal{H}_A)$ and $\dim(\mathcal{H}_B)$ are their respective dimensions.
\bd[Pure/Mixed States and Density Matrix]
 The \textbf{composite quantum system} $\mathcal{H}_{A}\ox \mathcal{H}_{B}$ is
\begin{itemize}
     \item A \textbf{pure state} is a vector $\ket{\psi}=\sum_{i}\ket{\psi}_i^A\ox\ket{\psi}_i^B$, where $\ket{\psi}_i^A$ and $\ket{\psi}_i^B$ are states in $\mathcal{H}_A$ and $\mathcal{H}_B$, respectively.
 	\item A \textbf{density matrix} for a pure state $\ket{\psi}$ is the matrix $\rho=\proj{\psi}$ in $M_{s}(\bbC)$, where $s=dim(\mathcal{H}_A)dim(\mathcal{H}_B)$.
 	\item A \textbf{mixed state} $\rho$ is the convex combination of density matrices of pure states $\{\psi_j\}_{j=1}^m$, $\rho=\sum_{j=1}^mp_j\proj{\psi_j}$, where $\sum_{j=1}^m{p_j}=1$ and $p_j>0$.
\end{itemize}
\ed

A pure state $\ket{\phi}_{AB}$ where $k=\dim(\mathcal{H}_A)$ and $l=\dim(\mathcal{H}_B)$ can be considered as a $k\times l$ matrix. For example, the pure entangled state $\ket{\phi}_{AB}=(1, 1, 0, 1)$ is equivalent to the matrix $\bma 1&1\\0&1 \ema$. Additionally, the set of all mixed states is exactly the cone of all positive semidefinite matrices.

\qquad


Dubbed ``spooky action at a distance" by Albert Einstein, \textbf{quantum entanglement} is a special connection between two quantum systems whereby the observation of one could instantaneously affect the other across an arbitrary distance. Two quantum systems that are not entangled are said to be separable. Quantum entanglement is more precisely defined with the following characteristics:

\bd[Entanglement]
A state $\rho$ in $\mathcal{H}_A\otimes\mathcal{H}_B$ is
\begin{itemize}
	\item \textbf{separable} if there exists $p_i>0$, $\rho^A_i$ and $\rho^B_i$ which are states in $\mathcal{H}_A$ and $\mathcal{H}_B$ such that $\rho=\sum_ip_i\rho^A_i\otimes\rho^B_i$, where $\sum_ip_i=1$.
	\item \textbf{entangled} if there is no such decomposition.
\end{itemize}
\ed

\bex[First Example of PPT Entanglement]
A separable state $\rho_0$ vs an entangled state $\rho_e$ in $M_3(\bbC)\ox M_3(\bbC)$. \cite{choi1982}
\bea\notag
\rho_0=\left[\begin{array}{ccc|ccc|ccc}
1 & 1 & 1 & 1 & 1 & 1 & 1 & 1 & 1 \\

1 & 1 & 1 & 1 & 1 & 1 & 1 & 1 & 1 \\

1 & 1 & 1 & 1 & 1 & 1 & 1 & 1 & 1 \\\hline

1 & 1 & 1 & 1 & 1 & 1 & 1 & 1 & 1 \\

1 & 1 & 1 & 1 & 1 & 1 & 1 & 1 & 1 \\

1 & 1 & 1 & 1 & 1 & 1 & 1 & 1 & 1 \\\hline

1 & 1 & 1 & 1 & 1 & 1 & 1 & 1 & 1 \\

1 & 1 & 1 & 1 & 1 & 1 & 1 & 1 & 1 \\

1 & 1 & 1 & 1 & 1 & 1 & 1 & 1 & 1 \\
\end{array}\right],\quad
\rho_e=\left[\begin{array}{ccc|ccc|ccc}
1 & \cdot & \cdot & \cdot & 1 & \cdot & \cdot & \cdot & 1 \\

\cdot & 2 & \cdot & 1 & \cdot & \cdot & \cdot & \cdot & \cdot \\

\cdot & \cdot & \frac{1}{2} & \cdot & \cdot & \cdot & 1 & \cdot & \cdot \\\hline

\cdot & 1 & \cdot & \frac{1}{2} & \cdot & \cdot & \cdot & \cdot & \cdot \\

1 & \cdot & \cdot & \cdot & 1 & \cdot & \cdot & \cdot & 1 \\

\cdot & \cdot & \cdot & \cdot & \cdot & 2 & \cdot & 1 & \cdot \\\hline

\cdot & \cdot & 1 & \cdot & \cdot & \cdot & 2 & \cdot & \cdot \\

\cdot & \cdot & \cdot & \cdot & \cdot & 1 & \cdot & \frac{1}{2} & \cdot \\

1 & \cdot & \cdot & \cdot & 1 & \cdot & \cdot & \cdot & 1 \\

\end{array}\right]
\eea
\eex

Determining whether an arbitrary quantum state is entangled is called the separability problem, and it has been proven to be NP-hard \cite{sevag2010}.

\vspace{2mm}

In this paper, linear algebra acts as the chief mathematical framework for analyzing quantum entanglement and quantum channels. Denote $M_n(\bbC)$ as the n-dimensional complex matrix algebra and $M_n^+(\bbC)$ be the set all positive semidefinite matrices in $M_n(\bbC)$. We consider a linear map $\phi$ between matrix algebras $M_n(\bbC)$ and $M_m(\bbC)$. In addition, denoted by $id_k$ and $\tau_k$ the identity and transpose map respectively on $M_k(\bbC)$.

\bd[Positivities]
A linear map $\phi$ from $M_n(\bbC)$ to $M_m(\bbC)$ is

\begin{itemize}
	\item \textbf{positive} if $\phi(M_n^+(\bbC))\subseteq M_m^+(\bbC)$.
	\item \textbf{$k$-positive} if $id_k\ox\phi$ is positive.
	\item \textbf{copositive} if $(\tau_m\circ\phi)(M_n^+(\bbC))\subseteq M_m^+(\bbC)$
	\item \textbf{$k$-copositive} if $\tau_k\ox\phi$ is positive.
	\item \textbf{completely positive} if it is $k$-positive for every $k$.
	\item \textbf{completely copositive} if it is $k$-copositive for every $k$.
	\item \textbf{decomposable} if it can be expressed as the sum of a completely positive map and a completely copositive map.
\end{itemize}
\ed
Let us explore and elaborate the definitions above through the following examples:
\bex[2-Positivity \& Transpose]
Although the transpose map is positive, it is not 2-positive. The transpose map 
\bea\notag
\tau_s: M_s\rightarrow M_s: x\mapsto x^t
\eea
Consider $\tau_2$ and look at the map
\bea\notag
id_2\ox\tau_2: M_2\ox M_2\rightarrow M_2\ox M_2.
\eea
It sends the positive semidefinite matrix to a matrix that is not positive semidefinite
\bea\notag
(id_2\ox\tau_2)\left(\bma
1 & 0 & 0 & 1\\
0 & 0 & 0 & 0\\
0 & 0 & 0 & 0\\
1 & 0 & 0 & 1\\
\ema\right)=
\bma
1 & 0 & 0 & 0\\
0 & 0 & 1 & 0\\
0 & 1 & 0 & 0\\
0 & 0 & 0 & 1\\
\ema
\eea
The transpose map is copositive. Similarity, $id_2$ is copositive but not 2-copositive.
\eex


A linear map can be represented by a matrix under the Choi-Jamiolkowski isomorphism. Such a matrix is called a Choi matrix, which is characterized by the following equation.
\bd[Choi Matrix]
Denote $e_{ij}$ the standard matrix units in $M_n(\bbC)$.
The Choi matrix of a linear map $\phi:M_n(\bbC)\rightarrow M_m(\bbC)$ is defined by
\bea\notag
C_{\phi}=\sum_{i,j=1}^n e_{ij}\ox \phi(e_{ij}).
\eea
\ed

\bex[Example of Choi Matrix]
A map $\psi:M_4\rightarrow M_4:x\mapsto tr(x)I_4-\frac{x}{2}$ represented in Choi matrix form is as follows:
\bea\notag
\large C_{\psi}=
\bma
\psi(e_{11}) & \psi(e_{12}) & \psi(e_{13}) & \psi(e_{14})\\
\psi(e_{21}) & \psi(e_{22}) & \psi(e_{23}) & \psi(e_{24})\\
\psi(e_{31}) & \psi(e_{32}) & \psi(e_{33}) & \psi(e_{34})\\
\psi(e_{41}) & \psi(e_{42}) & \psi(e_{43}) & \psi(e_{44})\ema=
\left[\begin{array}{cccc|cccc|cccc|cccc}
\frac{1}{2} & \cdot & \cdot & \cdot & \cdot & -\frac{1}{2} & \cdot & \cdot & \cdot & \cdot & -\frac{1}{2} & \cdot & \cdot & \cdot & \cdot & -\frac{1}{2} \\

\cdot & 1 & \cdot & \cdot & \cdot & \cdot & \cdot & \cdot & \cdot & \cdot & \cdot & \cdot & \cdot & \cdot & \cdot & \cdot \\

\cdot & \cdot & 1 & \cdot & \cdot & \cdot & \cdot & \cdot & \cdot & \cdot & \cdot & \cdot & \cdot & \cdot & \cdot & \cdot \\

\cdot & \cdot & \cdot & 1 & \cdot & \cdot & \cdot & \cdot & \cdot & \cdot & \cdot & \cdot & \cdot & \cdot & \cdot & \cdot \\\hline

\cdot & \cdot & \cdot & \cdot & 1 & \cdot & \cdot & \cdot & \cdot & \cdot & \cdot & \cdot & \cdot & \cdot & \cdot & \cdot \\

-\frac{1}{2} & \cdot & \cdot & \cdot & \cdot & \frac{1}{2} & \cdot & \cdot & \cdot & \cdot & -\frac{1}{2} & \cdot & \cdot & \cdot & \cdot & -\frac{1}{2} \\

\cdot & \cdot & \cdot & \cdot & \cdot & \cdot & 1 & \cdot & \cdot & \cdot & \cdot & \cdot & \cdot & \cdot & \cdot & \cdot \\

\cdot & \cdot & \cdot & \cdot & \cdot & \cdot & \cdot & 1 & \cdot & \cdot & \cdot & \cdot & \cdot & \cdot & \cdot & \cdot \\\hline

\cdot & \cdot & \cdot & \cdot & \cdot & \cdot & \cdot & \cdot & 1 & \cdot & \cdot & \cdot & \cdot & \cdot & \cdot & \cdot \\

\cdot & \cdot & \cdot & \cdot & \cdot & \cdot & \cdot & \cdot & \cdot & 1 & \cdot & \cdot & \cdot & \cdot & \cdot & \cdot \\

-\frac{1}{2} & \cdot & \cdot & \cdot & \cdot & -\frac{1}{2} & \cdot & \cdot & \cdot & \cdot & \frac{1}{2} & \cdot & \cdot & \cdot & \cdot & -\frac{1}{2} \\

\cdot & \cdot & \cdot & \cdot & \cdot & \cdot & \cdot & \cdot & \cdot & \cdot & \cdot & 1 & \cdot & \cdot & \cdot & \cdot \\\hline

\cdot & \cdot & \cdot & \cdot & \cdot & \cdot & \cdot & \cdot & \cdot & \cdot & \cdot & \cdot & 1 & \cdot & \cdot & \cdot \\

\cdot & \cdot & \cdot & \cdot & \cdot & \cdot & \cdot & \cdot & \cdot & \cdot & \cdot & \cdot & \cdot & 1 & \cdot & \cdot \\

\cdot & \cdot & \cdot & \cdot & \cdot & \cdot & \cdot & \cdot & \cdot & \cdot & \cdot & \cdot & \cdot & \cdot & 1 & \cdot \\

-\frac{1}{2} & \cdot & \cdot & \cdot & \cdot & -\frac{1}{2} & \cdot & \cdot & \cdot & \cdot & -\frac{1}{2} & \cdot & \cdot & \cdot & \cdot & \frac{1}{2} \\

\end{array}\right]
\eea
\eex


\bd[Partial Transpose]
Given a square matrix $A\ox B$, its partial transpose with respect to the first component is $A^{t}\ox B$. Similarly, its partial transpose with respect to the second component is $A\ox B^{t}$. Usually, $(A\ox B)^{\Gamma}$ denotes the partial transpose of $A\ox B$ with respect to the first component.
\ed
\bex[Partial Transpose]
Let A be a $3\times3$ matrix and B be a $2\times2$ matrix as follows:
\bea\notag
A=
\bma
1 & 2 & 1\\
2 & 1 & 2\\
1 & 2 & 1\\
\ema,
B=
\bma
0 & 1\\
-1 & 0\\
\ema
\eea
Hence
\bea\notag
A\ox B=
\left[\begin{array}{cc|cc|cc}
	0 & 1 & 0 & 2 & 0 & 1\\
	-1 & 0 & -2 & 0 & -1 & 0\\ \hline
	0 & 2 & 0 & 1 & 0 & 1\\
	-2 & 0 & -1 & 0 & -1 & 0\\ \hline
	0 & 1 & 0 & 2 & 0 & 1\\
	-1 & 0 & -2 & 0 & -1 & 0\\
\end{array}\right],
A^t\ox B=
\left[\begin{array}{cc|cc|cc}
	0 & 1 & 0 & 2 & 0 & 1\\
	-1 & 0 & -2 & 0 & -1 & 0\\ \hline
	0 & 2 & 0 & 1 & 0 & 1\\
	-2 & 0 & -1 & 0 & -1 & 0\\ \hline
	0 & 1 & 0 & 2 & 0 & 1\\
	-1 & 0 & -2 & 0 & -1 & 0\\
\end{array}\right],
A\ox B^{t}=
\left[\begin{array}{cc|cc|cc}
	0 & -1 & 0 & -2 & 0 & -1\\
	1 & 0 & 2 & 0 & 1 & 0\\ \hline
	0 & -2 & 0 & -1 & 0 & -1\\
	2 & 0 & 1 & 0 & 1 & 0\\ \hline
	0 & -1 & 0 & -2 & 0 & -1\\
	1 & 0 & 2 & 0 & 1 & 0\\
\end{array}\right].
\eea
\eex

\bpp[Linear Map vs. Choi matrix]
A map $\phi\in B(M_n(\bbC), M_m(\bbC))$ is completely positive iff its Choi matrix $C_{\phi}$ is positive semidefinite. Similarly, a map $\phi\in B(M_n(\bbC), M_m(\bbC))$ is completely copositive iff the partial transpose of its Choi matrix $C_{\phi}^{\Gamma}$ is positive semidefinite.
\epp

\bd[PPT]
A bipartite quantum state $\rho$ is said to be 
\begin{itemize}
	\item \textbf{positive partial transpose (PPT)} if the partial transpose with respect to the first system $\rho^{\Gamma_A}\ge0$ is still a PSD matrix
	\item \textbf{non-positive transpose (NPPT)} if $\rho^{\Gamma_A}$ has at least one negative eigenvalue.
\end{itemize}

\ed

The PPT test asks whether $\rho$ is PPT. Separability implies PPT but the converse is not always true. For $M_2(\bbC)\ox M_3(\bbC)$, all PPT states are separable \cite{peres1996}. However, in higher dimensions such as $M_3(\bbC)\ox M_3(\bbC)$ and $M_2(\bbC)\ox M_4(\bbC)$, there exist PPT states that are not separable \cite{choi1982,tang1986}. Searching for PPT entangled states in high dimensional quantum systems is an important task and has numerous applications in quantum communication. \cite{kye2013}

\vspace{2mm}
In the context of quantum information theory, quantum operations are implemented by quantum channels whose mathematical description is as follows \cite{choi1972}.

\bd[Quantum Channel]
A quantum channel is a completely positive trace-preserving (CPTP) linear map $\phi$ between matrix algebras $M_n(\bbC)$ and $M_m(\bbC)$. Furthermore, we call a quantum channel
\begin{itemize}
	\item \textbf{positive partial transpose (PPT)} if its Choi matrix $C_{\phi}$ is PPT.
	\item \textbf{entanglement breaking (EB)} if its Choi matrix $C_{\phi}$ is separable.
\end{itemize}

\ed

With the aforementioned definitions, we present the PPT Squared Conjecture \cite{report2012}. This Conjecture is included in the list of open problems posted in the website of Institute for Quantum Optics and Quantum Information (IQOQI) in Vienna compiled by Reinhard F. Werner and a team of researchers. The link is here \href{https://oqp.iqoqi.univie.ac.at/open-quantum-problems}{See problem 38}.

\bcj[PPT Squared Conjecture]
The PPT Square conjecture proposed by Matthias Christandl states that given a PPT quantum channel $\phi$ in $B(M_n(\bbC))$, the composite channel $\phi\circ\phi$ is an entanglement breaking channel. 
\ecj
Referencing figure 1, if $\dim(M_a) = \dim(M_b) = \dim(M_c)$ and $\phi_1 = \phi_2$, then the composite quantum channel $\phi_1\circ\phi_2$ will be entanglement breaking according to the conjecture.
In addition, this conjecture is dimension-dependent. It is proven to be valid for low dimensional cases ($n\leq 3$) \cite{chw2019,cyt2019}. This paper investigates whether there exists a counterexample in high dimensions. The general belief is that this kind of counterexample does exist.
\vspace{2mm}

\subsection{Recent Progress}
The conjecture has received a lot of attention recently. In the case $n\leq 2$, the conjecture becomes trivial, as shown in proposition 16. The conjecture only becomes meaningful when $n \geq 3$. It was recently proven that the conjecture holds true in dimension three and some examples such as the Gaussian quantum channels are proven to support the conjecture in all dimensions \cite{chw2019}. Another proof for the conjecture in the case $n = 3$ derives from the fact that every two-qutrit PPT states have Schmidt numbers that are at most two \cite{cyt2019}.

A noteworthy concept that is relevant to the PPT squared conjecture is the entanglement breaking index. The entanglement breaking index effectively measures the amount of noise introduced by a PPT quantum channel \cite{lg2015}.

\bd[Entanglement Breaking Index]
The \textbf{entanglement breaking index} $N$ is an integer-valued functional that measures the number of times a quantum channel $\phi$ needs to compose with itself in order to become entanglement-breaking (\textbf{EB}). It is mathematically characterized by the following equation:
\bea\notag
N(\phi)=min \{k \geq 1: \phi^k\ \textit{is entanglement-breaking}\}
\eea
\ed

Consider the Choi matrix of a PPT quantum channel $\phi\in B(M_2(\bbC))$. As every PPT state in $B(M_2(\bbC)\ox M_2(\bbC))$ is separable, $\phi$ is by default an entanglement-breaking channel. Recent progress shows that $N(\phi)\leq 2$ for every PPT quantum channel $\phi\in B(M_3(\bbC))$. Here's a brief summary:

\bpp[Schmidt Number in Low Dimensions]
$N(\phi)=1$ for $\phi\in B(M_n(\bbC))$ where $n \leq 2$. $N(\phi)=2$ is for $\phi\in B(M_3(\bbC))$.
\epp

The counterpart to an entanglement-breaking channel is an entanglement-saving channel. \cite{lg2016}
\bd[ES Channel]
An \textbf{entanglement saving (ES)} channel $\phi$ is a CPTP map that preserves the entanglement of a maximally entangled state after a finite, arbitrary iterations of repeated composition.
\ed

Within the set of ES channels, there exist two  important subsets - asymptotically entanglement saving (AES) channel and universal entanglement-preserving channel (UEP).
\bd[AES Channel]
An \textbf{asymptotically entanglement saving (AES)} quantum channel $\phi$ is a CPTP map whose entanglement breaking index is unbounded. That is, such a channel preserves entanglement even as the number of its composition approaches infinity.
\bea\notag
\lim_{n\to\infty} \phi^n \ \textit{is NOT entanglement-breaking}
\eea

\ed

\bd[UEP Channel]
A \textbf{universal entanglement-preserving channel} $\phi$ is a CPTP map that preserves the entanglement of any entangled state $\rho_{AB}$ regardless of how weak the entanglement is.
\ed

The distance between the repeated compositions of every unital or trace-preserving PPT channel and the set of entanglement breaking maps tends to zero \cite{kmp2017}. Furthermore, every unital PPT channel becomes entanglement breaking after a finite number of compositions \cite{rjp2018}. More generally, the notion of faithful quantum channels and its properties are is explored in \cite{hrf2019}\bd[Faithful Channel]
A \textbf{faithful quantum channel} is a quantum channel that preserves a full-rank state.
\ed

It has been proven that every faithful PPT quantum channel has a finite entanglement breaking index \cite{hrf2019}. A method to obtain the concrete bounds on the entanglement breaking index for any faithful quantum channel is also included.

As far as the authors know, no counter-example to the PPT Squared Conjecture in any dimension has been presented in the literature. Our goal for this note is to investigate possible methods for finding such a counterexample in higher dimensions.

\newpage
\section{Techniques and Methodologies}


\subsection{Quantum Entanglement Witness}

A classical approach to detect an entangled state is by using an entanglement witness $\psi$ to perform a paring with a quantum state $\rho$. The following definitions and propositions are from \cite{hhhh2009}.

\bd[Paring]
The \textbf{paring} between a quantum state $\rho\in M_{m}(\bbC)\ox M_n(\bbC)$ and a positive linear map $\psi\in B(M_m(\bbC),M_{n}(\bbC))$ is defined as
\bea\notag
\langle\rho,\psi\rangle=tr(\rho C_{\psi}^t).
\eea
\ed

\bpp[Separability Under Paring]
If the paring $\langle\rho,\psi\rangle$ is non-negative for every positive linear map $\psi$, then the state $\rho$ is separable. The converse is also true.
\epp

\bd[Entanglement Witness]
Given an entangled state $\rho$, there exists a positive linear map $\psi$ called an \textbf{entanglement witness} such that the paring
\bea\notag
\langle\rho,\psi\rangle=tr(\rho C_{\psi}^t) < 0
\eea
In this case, the linear map $\psi$ is said to detect the entangled state $\rho$.
\ed

We prove two useful propositions using our notations below.

\bpp[Entanglement Witness is Not Completely Positive]
A completely positive (CP) linear map cannot serve as an entanglement witness.
\epp
\bpf
Given an arbitrary quantum state $\rho_{AB}$, consider the paring $\langle\rho_{AB},\psi\rangle=tr(\rho_{AB}C_{\psi}^t)$ where $\psi$ is a completely positive map. Because $\rho_{AB}$ and $C_{\psi}$ are positive semidefinite matrices, $\rho_{AB}=SS^{\dagger}$ and $C_{\psi}^t=TT^{\dagger}$. We have the following
\bea\notag
tr(\rho_{AB}C_{\psi}^t) \xlongequal{\rho_{AB},C_{\psi} \text{ are PSD}}&tr(SS^{\dagger}TT^{\dagger})\\\notag
\xlongequal{tr(CD)=tr(DC)}&tr(S^{\dagger}TT^{\dagger}S)=tr((S^{\dagger}T)(S^{\dagger}T)^{\dagger})\\\notag
\xlongequal{C=S^{\dagger}T}&tr(CC^{\dagger})\geq0.\notag
\eea
\epf

\bpp[Indecomposable Entanglement Witness detects PPTES]
The entanglement witness of a PPT entangled state is an indecomposable positive linear map.
\epp
\bpf
The equivalence to the above proposition is that every decomposable positive linear maps as entanglement witnesses cannot detect PPT entangled states. Given an arbitrary PPT quantum state $\rho_{AB}$, consider the paring $\langle\rho_{AB},\psi\rangle=tr(\rho_{AB}C_{\psi}^t)$ where $\psi$ is a decomposable positive map. By the definition of decomposability, $\psi=\psi_1+\psi_2$ where $\psi_1$ is a completely positive linear map and $\psi_2$ is a completely co-positive linear map respectively. The partial transpose $\Gamma$ is taken with respect to the first subsystem $\mathcal{H}_{A}$. Therefore, we have the following:
\bea\notag
tr(\rho_{AB}C_{\psi}^t) \xlongequal{\psi=\psi_1+\psi_2}&tr(\rho_{AB}(C_{\psi_1}^t))+tr(\rho_{AB}C_{\psi_2}^t)\\\notag
\xlongequal{tr(D)=tr(D^{\Gamma_{A}})}&tr(\rho_{AB}(C_{\psi_1}^t))+tr((\rho_{AB}C_{\psi_2}^t)^{\Gamma})=tr(\rho_{AB}C_{\psi_1}^t)+tr(C_{\psi_2}^t)^{\Gamma}(\rho_{AB}^{\Gamma})\\\notag
\xlongequal{\rho_{AB} \text{ is PPT}}&tr(\rho_{AB}C_{\psi_1}^t)+tr(C_{\psi_2}^{\Gamma})^{t}(\rho_{AB}^{\Gamma})\geq0.\notag
\eea
The two addends $tr(\rho_{AB}C_{\psi_1}^t)$ and $tr(C_{\psi_2}^{\Gamma})^{t}(\rho_{AB}^{\Gamma})$ are nonnegative by the aforementioned proposition.
\epf

\subsection{Schmidt Rank and Schmidt Number}

The Schmidt rank and the Schmidt number are important notions that have been extensively used in the literature on quantum entanglement because it offers an elegant expression that illustrates the extent of entanglement of a bipartite quantum system \cite{sbl2001,bchhkls2002}. Let $\ket{\phi}_{AB}$ be an arbitrary vector in $\mathcal{H}_A\ox\mathcal{H}_B$.

\bd[Schmidt Decomposition]
The \textbf{Schmidt Decomposition} for
\begin{itemize}
	\item a bipartite pure state $\ket{\phi}_{AB}$ is
\bea\notag
\ket{\phi}_{AB} = \sum_{i, \mu}a_{i\mu}\ket{i}_{A}\ox\ket{\mu}_{B}.
\eea
Here $\{\ket{i}_A\}$ and $\{\ket{i}_B\}$ are the orthonormal basis for $\mathcal{H}_A$ and $\mathcal{H}_B$

	\item a bipartite mixed state $\ket{\rho}_{AB}$ is
\bea\notag
\rho_{AB}=\sum_{j}p_{j}\ket{\phi_j}_{AB}\bra{\phi_j}_{AB}
\eea
Here $\ket{\phi}_{AB}$ is a pure state in $\mathcal{H}_{A}\ox\mathcal{H}_{B}$ and $\sum_{j}p_j=1$
\end{itemize}

\ed

\bd[Schmidt Rank]
The \textbf{Schmidt Rank} $SR(\rho_{AB})$ for a pure state $\ket{\phi}_{AB}$ is defined by the rank of the corresponding matrix. 
\ed

\bd[Schmidt Number]
A bipartite density matrix $\rho_{AB}$ has \textbf{Schmidt Number} $SN(\rho_{AB})=k$ if 
\begin{itemize}
	\item for every Schmidt decomposition $\{p_j>0, \ket{\phi_j}_{AB}\}$ of $\rho_{AB}$,  at least one of the vectors $\ket{\phi_j}_{AB}$ has Schmidt rank at least $k$.
	\item there exists a decomposition of $\rho_{AB}$ with all vectors $\ket{\psi_i}$ of Schmidt rank at most $k$.
\end{itemize}
Equivalently, $SN(\rho)=\min\limits_{\rho=\sum_ip_i\ket{\phi_j}_{AB}\bra{\phi_j}_{AB}}\bigg\{\max\limits_{j} SR(\ket{\phi_j}_{AB})\bigg\}$. 
\ed

A quantum state $\rho$ is entangled iff it has a Schmidt number strictly greater than 1. Otherwise, $\rho$ is separable. The higher the Schmidt number is, the more entangled a state is. The Schmidt number of a state $\rho_{AB}\in M_m(\mathbb{C})\ox M_n(\mathbb{C})$ ranges from $1$ to $\min\{m, n\}$.

\subsection{Dual Cone}

Positive maps viewed as entanglement witnesses are classified by the following definitions and propositions \cite{kye2013}. There is a natural dual cone relation between the set of positive maps and the set of quantum states.

\bd[Various Sets]For all the quantum states in $M_m(\bbC)\ox M_n(\bbC)$, denote by
\begin{itemize}
	\item $P_k$ the set consisting all $k$-positive maps from $M_m(\bbC)$ to $M_n(\bbC)$.
	\item $V_k$ the set consisting all quantum states $\rho_{AB}$ such that $SN(\rho_{AB})\leq k$.
	\item $D$ the set consisting all decomposable maps from $M_m(\bbC)$ to $M_n(\bbC)$.
	\item $T$ the set consisting all PPT states.
\end{itemize}
\ed

It is natural to consider dual construction in convex geometry and that motivates the following definition.

\bd[Dual Pair]
A dual pair $(X, Y)$ under the bilinear paring $\langle\cdot, \cdot\rangle$ satisfies
\bea\notag
\forall \ x\in X \text{ and } \forall \ y\in Y, \text{ the paring } \langle x, y\rangle\geq0.
\eea
\ed
Hence $(D, T)$ is a dual pair. The following definition reveals the layers of entanglement witnesses. That is, nearly completely positive maps are less powerful in searching for entangled states. \cite{stm2010book} 

\bpp[Tower of Dual Pairs]
The sets of quantum states and positive linear maps sit in following tower ($m\leq n$).
\bea\notag
\begin{array}{ccccccccc}
	P_1 & \supseteq & P_2 & \supseteq & \cdots & \supseteq & P_m & \cong  & (M_m(\bbC)\ox M_n(\bbC))^+\\
	&&&&&&&& \parallel\\
	V_1 & \subseteq & V_2 & \subseteq & \cdots & \subseteq & V_m & = & (M_m(\bbC)\ox M_n(\bbC))^+\\
\end{array}
\eea
$(P_k,V_k)$ and $(D,T)$ are dual pairs under the bilinear paring $\langle\phi,\rho\rangle=tr(\rho C_{\phi}^t)$. The symbol $\cong$ stands for Choi-Jamiolkowski isomorphism between the set of completely positive maps and the set of positive semidefinite matrices.
\epp


\section{Schemes for Finding Counterexamples in High Dimensions}
Recall the PPT Squared Conjecture: If $\phi$ is a PPT channel on $M_n(\bbC)$, then $\phi\circ\phi$ is an entanglement-breaking channel. In the belief of the existence of a counterexample in high dimensions, we propose several schemes to search for it. Let us illustrate our schemes under the dimension $(n=4)$. Two facts to mention:
\begin{enumerate}
	\item If the composite channel $\phi\circ\phi$ is NOT entanglement breaking, then the map itself is NOT entanglement breaking. Hence we can start with a state as the Choi matrix of a PPT channel.
	\item The composite channel is PPT if the initial channel is PPT.
\end{enumerate}

\subsection{Most Naive Scheme}
A direct approach is to find out a PPT channel $\phi\in B(M_4(\bbC),M_4(\bbC))$ and then check if the composition channel $\phi\circ\phi$ is NOT entanglement breaking. That is, we have to find out the corresponding entanglement witness for the PPT entangled state $C_{\phi\circ\phi}$. Interestingly, few concrete examples of PPT channels between $M_4(\bbC)$ are presented in the literature. This makes the problem difficult to tackle through this scheme because most of the existing examples support the conjecture \cite{cyz2018}.

\begin{scheme}[Most Naive Scheme]\qquad

\begin{enumerate}[Step 1.]
\item Locate a PPT entangled state $\rho\in M_4(\bbC)\ox M_4(\bbC)$ as the Choi matrix of a channel $\phi\in B(M_4(\bbC),M_4(\bbC))$. 
	\item Write down the map $\phi$ from the chosen state $\rho$.
	\item Compute the composition $\phi\circ\phi$, check that the corresponding Choi matrix $C_{\phi\circ\phi}$ is entangled. 
\end{enumerate}
\end{scheme}

In step 1 \& 2 we have a PPT entangled state $\rho\in M_4(\bbC)\ox M_4(\bbC)$, then we think of it as the Choi matrix of a linear map $\phi\in B(M_4(\bbC),M_4(\bbC))$.
\bea\notag
\large C_{\phi}=
\bma
\phi(e_{11}) & \phi(e_{12}) & \phi(e_{13}) & \phi(e_{14})\\
\phi(e_{21}) & \phi(e_{22}) & \phi(e_{23}) & \phi(e_{24})\\
\phi(e_{31}) & \phi(e_{32}) & \phi(e_{33}) & \phi(e_{34})\\
\phi(e_{41}) & \phi(e_{42}) & \phi(e_{43}) & \phi(e_{44})\\
\ema
\eea

Hence the Choi matrix of $C_{\phi\circ\phi}$ is as follows.
\bea\notag
\large C_{\phi\circ\phi}=
\bma
(\phi\circ\phi)(e_{11}) & (\phi\circ\phi)(e_{12}) & (\phi\circ\phi)(e_{13}) & (\phi\circ\phi)(e_{14})\\
(\phi\circ\phi)(e_{21}) & (\phi\circ\phi)(e_{22}) & (\phi\circ\phi)(e_{23}) & (\phi\circ\phi)(e_{24})\\
(\phi\circ\phi)(e_{31}) & (\phi\circ\phi)(e_{32}) & (\phi\circ\phi)(e_{33}) & (\phi\circ\phi)(e_{34})\\
(\phi\circ\phi)(e_{41}) & (\phi\circ\phi)(e_{42}) & (\phi\circ\phi)(e_{43}) & (\phi\circ\phi)(e_{44})\ema\eea

In step 3, we have to find out an entanglement witness $\psi$ to verify that the state  $C_{\phi\circ\phi}$ is entangled. Here we include the tower of dual pairs when $n=4$.

\bd[Dual Pairs in $M_4(\bbC)$]
Consider all the quantum states in $M_4(\bbC)\ox M_4(\bbC)$. 
\begin{enumerate}
	\item Denote by $P_k$ the set consisting of all $k$-positive maps from $M_4(\bbC)$ to $M_4(\bbC)$.
	\item Denote by $V_k$ the set consisting of all quantum states $\rho$ whose Schmidt number is less than or equal to $k$.
	\item Denote by $D$ the set consisting of all decomposable maps from $M_4(\bbC)$ to $M_4(\bbC)$.
	\item Denote by $T$ the set consisting of all PPT entangled states.
\end{enumerate}
We have the following tower of sets.
\bea\notag
\begin{array}{ccccccccc}
	P_1 & \supseteq & P_2 & \supseteq & P_3 & \supseteq & P_4 & \cong  & (M_4(\bbC)\ox M_4(\bbC))^+\\
	&&&&&&&& \parallel\\
	V_1 & \subseteq & V_2 & \subseteq & V_3 & \subseteq & V_4 & = & (M_4(\bbC)\ox M_4(\bbC))^+\\
\end{array}
\eea
$(P_k,V_k)$ and $(D,T)$ are dual pairs under the bilinear paring $\langle\phi,\rho\rangle=tr(\rho C_{\phi}^t)$. 
\ed

To verify the PPT states $C_{\phi\circ\phi}\in M_4(\bbC)\ox M_4(\bbC)$ is of $\sn(\rho)=k>1$, we have to find out a corresponding entanglement witness in $P_{k-1}\backslash P_k$. This is extremely difficult since it is the genuine part of the separability problem. Hence we try to avoid it and move onto a revised scheme.

\subsection{Revised Scheme}

Bearing the difficulties of the aforementioned naive scheme in mind, we propose a revised scheme that tackles the problem from a different angle. First, we locate a PPT entangled state as the Choi matrix of the composite channel. Then we try to decompose it into two identical PPT channels. This saves us from verifying whether the composite channel is entanglement breaking or not.

\begin{scheme}[Revised Scheme A]\qquad
\begin{enumerate}[Step 1.]
	\item Consider an $M_4(\bbC)\ox M_4(\bbC)$ PPT Entangled state as the Choi matrix of a map $\Phi\in B(M_4(\bbC),M_4(\bbC))$
	\item Try to write $\Phi$ as a composition of the PPT maps $\phi$ with itself, where $\phi\in B(M_4(\bbC),M_4(\bbC))$.
\end{enumerate}
\end{scheme}

Let us abuse the usage of the terminology and speak of the channel $\phi$ as a square root of the channel $\Phi$. Such decomposition (or square rooting) of a PPTES could be carried out as a system of nonlinear equations, as shown in the following example.

\bex[Decomposition]\qquad

Here we present one method of decomposition. For the sake of simplicity, we will be using an NPPT state in $B(M_2(\bbC),M_2(\bbC))$. However, the logic applies equivalently to PPT states and any larger quantum systems.

Let $\phi$ be a quantum channel, expressed as a linear map whose entries are a linear composition of the input. There will be 16 coefficients in total, with the coefficients $a_{11}, a_{12}, a_{21}, a_{22}$ corresponding to  $x_{11}$, the coefficients $b_{11}, b_{12}, b_{21}, b_{22}$ corresponding to $x_{12}$, the coefficients $c_{11}, c_{12}, c_{21}, c_{22}$ corresponding to $x_{21}$, and the coefficients $d_{11}, d_{12}, d_{21}, d_{22}$ corresponding to $x_{22}$.
\bea\notag
\phi \left(\begin{array}{cc}
x_{11} & x_{12} \\
x_{21} & x_{22} \\
\end{array}\right)=
\left(\begin{array}{cc}
a_{11}x_{11}+b_{11}x_{12}+c_{11}x_{21}+d_{11}x_{22} & a_{12}x_{11}+b_{12}x_{12}+c_{12}x_{21}+d_{12}x_{22} \\
a_{21}x_{11}+b_{21}x_{12}+c_{21}x_{21}+d_{21}x_{22} & a_{22}x_{11}+b_{22}x_{12}+c_{22}x_{21}+d_{22}x_{22} \\
\end{array}\right)\eea

Denote $\psi$ composition $\phi \circ \phi$.

\bea\notag
\psi=\left(\begin{array}{cc}
y_{11} & y_{12} \\
y_{21} & y_{22} \\
\end{array}\right)\eea

where

\bea\notag
y_{11}= a_{11}(a_{11}x_{11}+b_{11}x_{12}+c_{11}x_{21}+d_{11}x_{22})\\\notag
+b_{11}(a_{12}x_{11}+b_{12}x_{12}+c_{12}x_{21}+d_{12}x_{22})\\\notag
+c_{11}(a_{21}x_{11}+b_{21}x_{12}+c_{21}x_{21}+d_{21}x_{22})\\\notag
+d_{11}(a_{22}x_{11}+b_{22}x_{12}+c_{22}x_{21}+d_{22}x_{22})\\\notag\\\notag
y_{12}=a_{12}(a_{11}x_{11}+b_{11}x_{12}+c_{11}x_{21}+d_{11}x_{22})\\\notag
+b_{12}(a_{12}x_{11}+b_{12}x_{12}+c_{12}x_{21}+d_{12}x_{22})\\\notag
+c_{12}(a_{21}x_{11}+b_{21}x_{12}+c_{21}x_{21}+d_{21}x_{22})\\\notag
+d_{12}(a_{22}x_{11}+b_{22}x_{12}+c_{22}x_{21}+d_{22}x_{22})\\\notag\\\notag
y_{21}=a_{21}(a_{11}x_{11}+b_{11}x_{12}+c_{11}x_{21}+d_{11}x_{22})\\\notag
+b_{21}(a_{12}x_{11}+b_{12}x_{12}+c_{12}x_{21}+d_{12}x_{22})\\\notag
+c_{21}(a_{21}x_{11}+b_{21}x_{12}+c_{21}x_{21}+d_{21}x_{22})\\\notag
+d_{21}(a_{22}x_{11}+b_{22}x_{12}+c_{22}x_{21}+d_{22}x_{22})\\\notag\\\notag
y_{22}=a_{22}(a_{11}x_{11}+b_{11}x_{12}+c_{11}x_{21}+d_{11}x_{22})\\\notag
+b_{22}(a_{12}x_{11}+b_{12}x_{12}+c_{12}x_{21}+d_{12}x_{22})\\\notag
+c_{22}(a_{21}x_{11}+b_{21}x_{12}+c_{21}x_{21}+d_{21}x_{22})\\\notag
+d_{22}(a_{22}x_{11}+b_{22}x_{12}+c_{22}x_{21}+d_{22}x_{22})\\\notag
\eea\notag

The Choi matrix of $\psi$, $C_{\psi}$ hence becomes

\scalebox{0.85}{%
  \begin{minipage}{0.0\linewidth}
\bea\notag \left[\begin{array}{cc|cc}
a_{11}^2+a_{12}b_{11}+a_{21}c_{11}+a_{22}d_{11} & a_{11}a_{12}+a_{12}b_{12}+a_{21}c_{12}+a_{22}d_{12} & a_{11}b_{11}+b_{11}b_{12}+b_{21}c_{11}+b_{22}d_{11} & b_{12}^2+a_{12}b_{11}+b_{21}c_{12}+b_{22}d_{12} \\

a_{11}a_{21}+a_{12}b_{21}+a_{21}c_{21}+a_{22}d_{21} & a_{11}a_{22}+a_{12}b_{22}+a_{21}c_{22}+a_{22}d_{22} & a_{21}b_{11}+b_{12}b_{21}+b_{21}c_{21}+b_{22}d_{21} & a_{22}b_{11}+b_{12}b_{22}+b_{21}c_{22}+b_{22}d_{22} \\\hline

a_{11}c_{11}+b_{11}c_{12}+c_{11}c_{21}+c_{22}d_{11} & a_{12}c_{11}+b_{12}c_{12}+c_{12}c_{21}+c_{22}d_{12} & a_{11}d_{11}+b_{11}d_{12}+c_{11}d_{21}+d_{11}d_{22} & a_{12}d_{11}+b_{12}d_{12}+c_{12}d_{21}+d_{12}d_{22} \\

c_{21}^2+a_{21}c_{11}+b_{21}c_{12}+c_{22}d_{21} & a_{22}c_{11}+b_{22}c_{12}+c_{21}c_{22}+c_{22}d_{22} & a_{21}d_{11}+b_{21}d_{12}+c_{21}d_{21}+d_{21}d_{22} & d_{22}^2+a_{22}d_{11}+b_{22}d_{12}+c_{22}d_{21} \\

\end{array}\right]
\eea
\end{minipage}
}
\\
\vspace{2mm}

Now, if we are given the Choi matrix of $\psi$, which is what we usually encounter in the literature, we can use the 16 entries of $C_{\psi}$ to generate a system of equations and solve for the 16 variables. 

For example, given a state $\eta$ whose Choi matrix
\bea\notag
C_{\eta} =
\left(\begin{array}{cc|cc}
\frac{17}{4} & \cdot & \cdot & \frac{1}{9} \\
\cdot & 2 & \cdot & \cdot \\\hline
\cdot & \cdot & 2 & \cdot \\
\frac{1}{9} & \cdot & \cdot & \frac{17}{4} \\
\end{array}\right)\eea

We can match each of the numerical entry of $C_{\eta}$ to each of the expressions of the entries of $C_{\psi}$, thus generating 16 systems of equation. If we use MATLAB's solve function, we could actually acquire the decomposition of this matrix. $\eta=\phi \circ \phi$ where 
\bea\notag
\phi =
\left(\begin{array}{cc}
2x_{11}+\frac{1}{2}x_{22} & \frac{1}{3}x_{12} \\
\frac{1}{3}x_{21} & 2x_{22}+\frac{1}{2}x_{11} \\
\end{array}\right)\eea
\eex

The same logic could be applied to a PPTES in $B(M_4(\bbC),M_4(\bbC))$. Except in this scenario, there would be 256 (16 coefficients per entry $\times$ 16 entries in the Choi matrix) coefficients and 256 equations. Most of these equations, upon simplification, could yield 0 on both sides and hence could be eliminated. Therefore, usually we can simplify it down to around 40 equations and 40 variables (Note: the number of variables must be less or equal to the number of equations or we may not get a result). We have attached a Microsoft Excel spreadsheet to the supplementary documents of the submission that simplifies the system of equations.

It is important to note that this square root may not be unique, as shown by the following example.
\bex[Non-uniqueness]
Consider the following completely positive linear map $\psi\in B(M_2(\bbC),M_2(\bbC))$

\bea\notag \psi\bigg(\left[\begin{array}{cc}
x_{11} & x_{12} \\

x_{21} & x_{22} \\
\end{array}\right]\bigg)= \left[\begin{array}{cc}
4x_{11} & x_{12} \\

x_{21} & 4x_{22} \\
\end{array}\right]
\eea

Such a map can be decomposed (or taken square root of) in several ways. In other words, $\phi(X)$ where $\phi\circ \phi=\psi$ can be any of the following:

\bea\notag
\left[\begin{array}{cc}
2x_{11} & x_{12} \\

x_{21} & 2x_{22} \\

\end{array}\right],
\left[\begin{array}{cc}
2x_{11} & x_{21} \\

x_{12} & 2x_{22} \\

\end{array}\right],
\left[\begin{array}{cc}
2x_{11} & -x_{12} \\

-x_{21} & 2x_{22} \\

\end{array}\right],
\left[\begin{array}{cc}
2x_{11} & -x_{21} \\

-x_{12} & 2x_{22} \\

\end{array}\right]
\eea
Therefore, upon finding a decomposition, it is necessary to verify if the resultant is a positive and PPT state (we have also produced a Python program to do so. See the Appendix).
\eex

\br[Strengths and Limitations of Revised Scheme A]
This approach is mathematical straightforward. With a powerful enough computer, such a decomposition could be calculated. In addition, there are only two constraints this decomposition needs to meet: 1) the resulting state has to be positive. 2) the resulting state is a PPT state. Nevertheless, there exist several limitations to this approach.
\begin{itemize}
	\item Even though we can reduce a 256-equation decomposition down to around 40 variables, this quantity still appears to be too much for MATLAB on our computer to handle.
	\item The system of equations generated through this decomposition is NOT linear.
\end{itemize}
\er

\begin{scheme}[Revised Scheme B]\qquad
\begin{enumerate}[Step 1.]
	\item Consider a $M_4(\bbC)\ox M_4(\bbC)$ PPT entangled state as the Choi matrix of a map $\psi\in B(M_4(\bbC), M_4(\bbC))$
	\item Try to write $\psi$ as a composition of two PPT maps $\phi_1\in B(M_4(\bbC), M_b(\bbC))$ and $\phi_2\in B(M_b(\bbC), M_4(\bbC))$ where $b$ can be any positive integer.	
	\item Check whether Choi matrixes $\phi_1$ and $\phi_2$ are PPT quantum channels.
\end{enumerate}
\end{scheme}

To understand why such a scheme works, we need to consider the following. In physics, for quantum channel $\phi$, the dimension of the domain and the dimension of the codomain mentioned in the PPT squared conjecture are the same. But if we treat the conjecture as a purely mathematical problem, we are able to reformulate it in the following question.
\begin{question}[Modification on Dimensions]\quad\\
	If $\phi_1\in B(M_a(\bbC),M_b(\bbC))$ and $\phi_2\in B(M_b(\bbC),M_c(\bbC))$ are completely positive and completely copositive maps, then is the Choi matrix of composition map $C_{\phi_2\circ\phi_1}$ is of Schmidt number one?
\end{question}

Note that the question is equivalent to the original PPT squared conjecture after a dimension modification \cite{chw2019}. In their proof, a counterexample of the PPT squared conjecture for $n=2\max\{a,b,c\}$ can be obtained via a counterexample in the above question.

\begin{scheme}[Revised Scheme C]\qquad\\
The base case $a=b=c=3$ is answered affirmatively, but any rise in either $a$, $b$ or $c$ will leave the question open. Hence the following scheme is one of the modifications next to consider. We raise the middle index $b$ and keep the other two indexes to begin with.
\begin{enumerate}[Step 1.]
	\item Consider $M_3(\bbC)\ox M_3(\bbC)$ PPT entangled states as Choi matrix of a map $\psi\in B(M_3(\bbC), M_3(\bbC))$
	\item Try to write $\psi$ as a composition of two PPT maps $\phi_1\in B(M_3(\bbC), M_4(\bbC))$ and $\phi_2\in B(M_4(\bbC), M_3(\bbC))$.	
	\item Check the Choi matrixes $\phi_1$ and $\phi_2$ are PPT quantum channels.
\end{enumerate}
\end{scheme}

\bea\notag C(\phi)=\left[\begin{array}{ccc|ccc|ccc}
1 & \cdot & \cdot & \cdot & 1 & \cdot & \cdot & \cdot & 1 \\

\cdot & 2 & \cdot & 1 & \cdot & \cdot & \cdot & \cdot & \cdot \\

\cdot & \cdot & \frac{1}{2} & \cdot & \cdot & \cdot & 1 & \cdot & \cdot \\\hline

\cdot & 1 & \cdot & \frac{1}{2} & \cdot & \cdot & \cdot & \cdot & \cdot \\

1 & \cdot & \cdot & \cdot & 1 & \cdot & \cdot & \cdot & 1 \\

\cdot & \cdot & \cdot & \cdot & \cdot & 2 & \cdot & 1 & \cdot \\\hline

\cdot & \cdot & 1 & \cdot & \cdot & \cdot & 2 & \cdot & \cdot \\

\cdot & \cdot & \cdot & \cdot & \cdot & 1 & \cdot & \frac{1}{2} & \cdot \\

1 & \cdot & \cdot & \cdot & 1 & \cdot & \cdot & \cdot & 1 \\

\end{array}\right]
\eea

Hence the corresponding map $\psi$ writes:

\bea\notag \psi\left(\begin{array}{ccc}
x_{11} & x_{12} & x_{13} \\

x_{21} & x_{22} & x_{23} \\

x_{31} & x_{32} & x_{33} \\

\end{array}\right)
= 
\left(\begin{array}{ccc}
x_{11}+\frac{1}{2}x_{22}+2x_{33} & x_{12}+x_{21} & x_{13}+x_{31} \\

x_{12}+x_{21} & 2x_{11}+x_{22}+\frac{1}{2}x_{33} & x_{23}+x_{32} \\

x_{13}+x_{31} & x_{23}+x_{32} & \frac{1}{2}x_{11}+2x_{22}+x_{33} \\

\end{array}\right)
\eea

It is unknown whether the decomposition is possible or not. The main difficulty is the huge number of variables to determine when decomposing the channel. In addition, the fact that 4 is an even number and 3 is an odd number also makes the decomposition harder.

\begin{scheme}[Revised Scheme D: Decomposition]\qquad\\
Adjusting the triplet of indices $(a,b,c)$ to be $(4,2,4)$ yields yet another unanswered formulation of the problem and the corresponding scheme is as follows. We believe it is the most promising scheme. 
\begin{enumerate}[Step 1.]
	\item Consider $M_4(\bbC)\ox M_4(\bbC)$ PPT entangled states as Choi matrix of a map $\Psi\in B(M_4(\bbC), M_4(\bbC))$
	\item Try to write $\psi$ as a composition of two PPT maps $\phi_1\in B(M_4(\bbC), M_2(\bbC))$ and $\phi_2\in B(M_2(\bbC), M_4(\bbC))$.
	\item Check the Choi matrixes $\phi_1$ and $\phi_2$ are PPT quantum channels.
\end{enumerate}
\end{scheme}

This scheme is similar to the previous scheme, the advantage is that the dimension of the middle system equals two. The fact that both two and four are even numbers also make the decomposition substantially easier. That reduces the complexity in determining the variables in the process of decomposing the channel.

Shifting from the decomposition point of view to the composition point of view yields the following scheme.

\begin{scheme}[Revised Scheme D: Composition]\qquad
\begin{enumerate}[Step 1.]
	\item Find a PPTES $\rho_1\in M_4(\bbC)\ox M_2(\bbC)$ as the Choi matrix of a channel $\phi_1\in B(M_4(\bbC),M_2(\bbC))$. 
	\item Find another PPTES $\rho_2\in M_2(\bbC)\ox M_4(\bbC)$ as the Choi matrix of a channel $\phi_2\in B(M_2(\bbC),M_4(\bbC))$. 	
	\item Write down the map $\phi_1$ and $\phi_2$.
	\item Compute the composition $\phi_1\circ\phi_2$, check that the corresponding Choi matrix $C_{\phi\circ\phi}$ is entangled. 
\end{enumerate}
\end{scheme}

Let us generate two PPT quantum channels using a concrete PPTES from \cite{agkl2010}, compute their composition and try to check whether the composite channel is entanglement breaking in the next example.

\bex[Possible Counterexample for $(a,b,c)=(4,2,4)$]\notag
\qquad\\
Consider the following $8\times8$ positive definite matrix $\rho$ and its partial transpose with respect to $4\ox2$ system and $2\ox4$ system, respectively.
\bea\notag \rho=\left[\begin{array}{cc|cc|cc|cc}
a & \cdot & \cdot & \cdot & \cdot & \cdot & \cdot & t \\

\cdot & 1 & \cdot & \cdot & -1 & \cdot & \cdot & \cdot \\\hline

\cdot & \cdot & \frac{1}{a} & \cdot & \cdot & -1 & \cdot & \cdot \\

\cdot & \cdot & \cdot & 1 & \cdot & \cdot & -1 & \cdot \\\hline

\cdot & -1 & \cdot & \cdot & 1 & \cdot & \cdot & \cdot \\

\cdot & \cdot & -1 & \cdot & \cdot & \frac{1}{a} & \cdot & \cdot \\\hline

\cdot & \cdot & \cdot & -1 & \cdot & \cdot & 1 & \cdot \\

t & \cdot & \cdot & \cdot & \cdot & \cdot & \cdot & a \\
\end{array}\right]
=\left[\begin{array}{cccc|cccc}
a & \cdot & \cdot & \cdot & \cdot & \cdot & \cdot & t \\

\cdot & 1 & \cdot & \cdot & -1 & \cdot & \cdot & \cdot \\

\cdot & \cdot & \frac{1}{a} & \cdot & \cdot & -1 & \cdot & \cdot \\

\cdot & \cdot & \cdot & 1 & \cdot & \cdot & -1 & \cdot \\\hline

\cdot & -1 & \cdot & \cdot & 1 & \cdot & \cdot & \cdot \\

\cdot & \cdot & -1 & \cdot & \cdot & \frac{1}{a} & \cdot & \cdot \\

\cdot & \cdot & \cdot & -1 & \cdot & \cdot & 1 & \cdot \\

t & \cdot & \cdot & \cdot & \cdot & \cdot & \cdot & a \\

\end{array}\right]
\eea

The partial transposes w.r.t. $4\ox2$ and $2\ox4$ are
\bea\notag
\rho^{\Gamma_A}=\left[\begin{array}{cc|cc|cc|cc}
a & \cdot & \cdot & \cdot & \cdot & -1 & \cdot & \cdot \\

\cdot & 1 & \cdot & \cdot & \cdot & \cdot & t & \cdot \\\hline

\cdot & \cdot & \frac{1}{a} & \cdot & \cdot & \cdot & \cdot & -1 \\

\cdot & \cdot & \cdot & 1 & -1 & \cdot & \cdot & \cdot \\\hline

\cdot & \cdot & \cdot & -1 & 1 & \cdot & \cdot & \cdot \\

-1 & \cdot & \cdot & \cdot & \cdot & \frac{1}{a} & \cdot & \cdot \\\hline

\cdot & t & \cdot & \cdot & \cdot & \cdot & 1 & \cdot \\

\cdot & \cdot & -1 & \cdot & \cdot & \cdot & \cdot & a \\
\end{array}\right]
\quad \rho^{\Gamma_A}=
\left[\begin{array}{cccc|cccc}
a & \cdot & \cdot & \cdot & \cdot & -1 & \cdot & \cdot \\

\cdot & 1 & \cdot & \cdot & \cdot & \cdot & -1 & \cdot \\

\cdot & \cdot & \frac{1}{a} & \cdot & \cdot & \cdot & \cdot & -1\\

\cdot & \cdot & \cdot & 1 & t & \cdot & \cdot & \cdot \\\hline

\cdot & \cdot & \cdot & t & 1 & \cdot & \cdot & \cdot \\

-1 & \cdot & \cdot & \cdot & \cdot & \frac{1}{a} & \cdot & \cdot \\

\cdot & -1 & \cdot & \cdot & \cdot & \cdot & 1 & \cdot \\

\cdot & \cdot & -1 & \cdot & \cdot & \cdot & \cdot & a \\
\end{array}\right]
\eea
Direct computation shows when $0<a<1$ and $|t|<a$, the state $\rho$ is PPTES in $2\ox4$ and PPT in $4\ox2$ \cite{agkl2010}. We do not know whether the state $\rho$ entangled in $4\ox2$ or not.


The induced maps $\phi_1$ and $\phi_2$ are as follows.
\bea\notag
\phi_1
\left(\begin{array}{cccc}
x_{11} & x_{12} & x_{13} & x_{14} \\
x_{21} & x_{22} & x_{23} & x_{24} \\
x_{31} & x_{32} & x_{33} & x_{34} \\
x_{41} & x_{42} & x_{43} & x_{44} \\
\end{array}\right)
= \left(\begin{array}{cc}
ax_{11}+\frac{1}{a}x_{22}+x_{33}+x_{44} & tx_{41}-x_{32}-x_{24}-x_{13} \\
tx_{14}-x_{23}-x_{31}-x_{42} & ax_{44}+\frac{1}{a}x_{33}+x_{11}+x_{22} \\
\end{array}\right)\eea
\bea\notag
\phi_2 \left(\begin{array}{cc}
y_{11} & y_{12} \\
y_{21} & y_{22} \\
\end{array}\right)=
\left(\begin{array}{cccc}
ay_{11}+y_{22} & -y_{12} & \cdot & \cdot \\
-y_{21} & y_{11}+\frac{1}{a}y_{22} & -y_{12} & \cdot \\
\cdot & -y_{21} & \frac{1}{a}y_{11}+y_{22} & -y_{12} \\
\cdot & \cdot & -y_{21} & y_{11}+ay_{22} \\
\end{array}\right)\eea

Hence the composition $\phi_2\circ\phi_1$ is
\bea\notag
(\phi_2\circ\phi_1)\left(\begin{array}{cccc}
x_{11} & x_{12} & x_{13} & x_{14} \\
x_{21} & x_{22} & x_{23} & x_{24} \\
x_{31} & x_{32} & x_{33} & x_{34} \\
x_{41} & x_{42} & x_{43} & x_{44} \\
\end{array}\right)\eea

\scalebox{0.7}{%
  \begin{minipage}{0.0\linewidth}
\bea\notag
=\left(\begin{array}{cccc}
(a^2+1)x_{11}+2x_{22}+(a+\frac{1}{a})x_{33}+2ax_{44} & -tx_{41}+x_{32}+x_{24}+x_{13} & \cdot & \cdot \\
-tx_{14}+x_{23}+x_{31}+x_{42} & (a+\frac{1}{a})x_{11}+\frac{2}{a}x_{22}+(1+\frac{1}{a^2})x_{33}+2x_{44} & -tx_{41}+x_{32}+x_{24}+x_{13} & \cdot \\
\cdot & -tx_{14}+x_{23}+x_{31}+x_{42} & 2x_{11}+(1+\frac{1}{a^2})x_{22}+\frac{2}{a}x_{33}+(a+\frac{1}{a})x_{44} & -tx_{41}+x_{32}+x_{24}+x_{13} \\
\cdot & \cdot & -tx_{14}+x_{23}+x_{31}+x_{42} & 2ax_{11}+(a+\frac{1}{a})x_{22}+2x_{33}+(a^2+1)x_{44} \\
\end{array}\right)\eea	
\end{minipage}
}
\\
\vspace{2mm}

The corresponding Choi matrix of the composite channel is:
\bea\notag
C_{\phi_2\circ\phi_1}=\left[\begin{array}{cccc|cccc|cccc|cccc}
a^2+1 & \cdot & \cdot & \cdot & \cdot & \cdot & \cdot & \cdot & \cdot & 1 & \cdot & \cdot & \cdot & \cdot & \cdot & \cdot \\

\cdot & a+\frac{1}{a} & \cdot & \cdot & \cdot & \cdot & \cdot & \cdot & \cdot & \cdot & 1 & \cdot & -t & \cdot & \cdot & \cdot \\

\cdot & \cdot & 2 & \cdot & \cdot & \cdot & \cdot & \cdot & \cdot & \cdot & \cdot & 1 & \cdot & -t & \cdot & \cdot \\

\cdot & \cdot & \cdot & 2a & \cdot & \cdot & \cdot & \cdot & \cdot & \cdot & \cdot & \cdot & \cdot & \cdot & -t & \cdot \\\hline

\cdot & \cdot & \cdot & \cdot & 2 & \cdot & \cdot & \cdot & \cdot & \cdot & \cdot & \cdot & \cdot & 1 & \cdot & \cdot \\

\cdot & \cdot & \cdot & \cdot & \cdot & \frac{2}{a} & \cdot & \cdot & 1 & \cdot & \cdot & \cdot & \cdot & \cdot & 1 & \cdot \\

\cdot & \cdot & \cdot & \cdot & \cdot & \cdot & 1+\frac{1}{a} & \cdot & \cdot & 1 & \cdot & \cdot & \cdot & \cdot & \cdot & 1 \\

\cdot & \cdot & \cdot & \cdot & \cdot & \cdot & \cdot & a+\frac{1}{a} & \cdot & \cdot & 1 & \cdot & \cdot & \cdot & \cdot & \cdot \\\hline

\cdot & \cdot & \cdot & \cdot & \cdot & 1 & \cdot & \cdot & a+\frac{1}{a} & \cdot & \cdot & \cdot & \cdot & \cdot & \cdot & \cdot \\

1 & \cdot & \cdot & \cdot & \cdot & \cdot & 1 & \cdot & \cdot & 1+\frac{1}{a^2} & \cdot & \cdot & \cdot & \cdot & \cdot & \cdot \\

\cdot & 1 & \cdot & \cdot & \cdot & \cdot & \cdot & 1 & \cdot & \cdot & \frac{2}{a} & \cdot & \cdot & \cdot & \cdot & \cdot \\

\cdot & \cdot & 1 & \cdot & \cdot & \cdot & \cdot & \cdot & \cdot & \cdot & \cdot & 2 & \cdot & \cdot & \cdot & \cdot \\\hline

\cdot & -t & \cdot & \cdot & \cdot & \cdot & \cdot & \cdot & \cdot & \cdot & \cdot & \cdot & 2a & \cdot & \cdot & \cdot \\

\cdot & \cdot & -t & \cdot & 1 & \cdot & \cdot & \cdot & \cdot & \cdot & \cdot & \cdot & \cdot & 2 & \cdot & \cdot \\

\cdot & \cdot & \cdot & -t & \cdot & 1 & \cdot & \cdot & \cdot & \cdot & \cdot & \cdot & \cdot & \cdot & a+\frac{1}{a} & \cdot \\

\cdot & \cdot & \cdot & \cdot & \cdot & \cdot & 1 & \cdot & \cdot & \cdot & \cdot & \cdot & \cdot & \cdot & \cdot & 1+\frac{1}{a^2} \\

\end{array}\right]
\eea
\eex

According to Mathematica(or see Appendix), we know that $C_{\phi_2\circ \phi_1}$ is always a full rank PPT state given the aforementioned range of $a$ and $t$. As a result, it is difficult to check whether it is entangled. Most of the existing entanglement witnesses in the literature cannot detect it. Nevertheless, it has a high chance of being an entangled state as the vast majority of states are entangled. Our hope is that it is indeed entangled and thus offers us the desired counterexample for the conjecture.

It is noteworthy that indecomposable 1-positive maps are relatively powerful entanglement witnesses to perform the checking according to the tower of sets. However, the entanglement witnesses are organized in a tree-structure, rather than a linear relationship, so two branches have different incomparable maximals. Hence it makes paring a PPTES with the right entanglement witness difficult.

\section{Summary}
In the first chapter, starting from the physics motivation of the conjecture, we have included the necessary fundamental knowledge of linear algebra and quantum information thereby producing a self-contained note. We then mentioned several of the recent progress. In the second chapter, we introduced concepts highly relevant to the conjecture such as a quantum measure, Schmidt number, the structure of quantum states, and positive maps. In the third chapter, we have developed two main approaches to attack the problem, the first being a decomposition of PPT quantum channels and the second being a composition of PPT quantum channels in unsolved dimensions. From these two approaches we devised numerous schemes. The decomposition scheme is hard to accomplish due to the number of variables and the nonlinearity of its system of equations. The composition scheme yields a potential counterexample.

\section{Appendix}
Along the way of our research, we have developed several useful pieces of programs for computation and writing.
\subsection{Localhost Website Produced With HTML, CSS, and JavaScript For Automating the Conversion of a Matrix to MATLAB and LaTEX Format}

\lstset{language=HTML}
\begin{lstlisting}[caption=index.html]

<!DOCTYPE html>
<html>
    <head>
        <title>Matrix Formatter v1</title>
        <script src="jquery.js"></script>
    	<script src="script.js"></script>
    	<style>
    		.yolo {
    			background-color: black;
    			padding: 5px;
			    color: white;
			    display: inline-block;
    		    margin-top: 10px;
    		    cursor: pointer;
		    }

		    .matrix-input {
		        margin: 10px;
		        width: 250px;
		        height: 20px;
		    }

		    #variables-input {
		        width: 80%;
		    }

		    #variables {
		        display: none;
		    }
    	</style>
    </head>
    <body>
        <div class="matrix-attr">
            <form>
                # of rows:<br>
                <input type="text" name="rows" id="rows"><br>
                # of columns:<br>
                <input type="text" name="columns" id="columns"><br>
                <span id="info" class="yolo">Submit</span>
            </form>
            <h1>Format Guidelines</h1>
            <ol>
                <li>Do not add extra spaces in your expression</li>
                <li>Write the multiplication signs</li>
                <li>Write subscript with _{subscript}, so x subscript 12 would be x_{12}</li>
                <li>In latex, some math functions like sqrt use {} instead of (). However, if you enter sqrt(), you have to change it manually in latex</li>
                <li>You MAY enter \</li>
                <li><strong>If you are using a fraction, you may enter them as a/b as long as both a and b are  numbers. If either the denominator or numerator contain something other than numbers, like x_{12}, please enter it with a parenthesis as (x_{12}) </strong></li>
                <li>For symbol compatibility for MATLAB, we change \ to s, { and } to _</li>
                <li><strong>You MAY NOT enter a space into the matrix entries</strong></li>
            </ol>
            <form id="variables">
                <h2>Please enter all variables, separated by commas. This is for helping you create symbolic functions in MATLAB</h2>
                <p>E.g. you may enter:\mu,x_{12},y</p>
                
                <input type="text" name="variables" id="variables-input"><br>
                <h1>Below this, please enter the entries of your matrix</h1>
            </form>
            <form id="matrix">
                
            </form>
            <div class="result">
            </div>
        </div>
    </body>
    
</html>

\end{lstlisting}
\lstset{language=JavaScript}
\begin{lstlisting}[caption=script.js]
var rows = 0;
var columns = 0;
var screenHeight = $(window).height();
var checkNum = function(c) {
    return ('0123456789'.indexOf(c) !== -1)
};

var matlabFriendly = function(text) {
    while(text.indexOf("{") != -1) {
        text = text.replace("{", "_")
    }
    while(text.indexOf("}") != -1) {
        text = text.replace("}", "_")
    }
    while(text.indexOf("\\") != -1) {
        text = text.replace("\\", "s")
    }
    return text;
}

var convertAlgebraToLATEX = function(expression) {
    if (expression == "0") {
        return "\\cdot"
    }
    var indexOfFraction = expression.indexOf("/")
    while (indexOfFraction != -1) {

        var numerator = expression[indexOfFraction-1];
        var numeratorStart = indexOfFraction - 1
        
        if (numerator == ")") {
            countOfParenthesis = 1
            numeratorStart = indexOfFraction - 2
            while(countOfParenthesis > 0) {
                
                do {
                    numerator = expression[numeratorStart] + numerator
                    if (expression[numeratorStart] == ")") {
                        countOfParenthesis++;
                        console.log("here")
                        console.log(countOfParenthesis)
                    }
                    numeratorStart--;
                } while (expression[numeratorStart] != "(");
                countOfParenthesis--;
                console.log(countOfParenthesis)
            }
        
            
        } else if (checkNum(numerator)) {
            numeratorStart = indexOfFraction - 2
            while(checkNum(expression[numeratorStart])) {
                numerator = expression[numeratorStart] + numerator
                numeratorStart--;
            }
        } else {
            alert("Failure occuring at " + expression[indexOfFraction - 3, indexOfFraction + 3]);
        }
        numeratorStart ++;
        console.log("Numerator start is")
        console.log(numeratorStart)
        
        
        // Denominator
        var denominator = expression[indexOfFraction+1];
        var endofDenominator = indexOfFraction + 1
        
        if (denominator == "(") {
            countOfParenthesis = 1
            
            endofDenominator = indexOfFraction + 2
            denominator = ""
            while(countOfParenthesis > 0) {
                
                do {
                    denominator += expression[endofDenominator]
                    if (expression[endofDenominator] == "(") {
                        countOfParenthesis++;
                    }
                    endofDenominator++;
                } while (expression[endofDenominator] != ")");
                countOfParenthesis--;
                
            }
            endofDenominator++;
        } else if (checkNum(denominator)) {
            endofDenominator = indexOfFraction + 2
            while(checkNum(expression[endofDenominator])) {
                denominator = expression[endofDenominator] + denominator
                endofDenominator++;
            }
            
        } else {
            alert("Failure occuring at " + expression[indexOfFraction - 3, indexOfFraction + 3]);
        }
        
        var replaceString = "\\frac{" + numerator + "}{" + denominator + "}"
        var toBeReplaced = expression.substring(numeratorStart, endofDenominator)
        console.log("to be replaced is ")
        console.log(toBeReplaced)
        console.log("replacee")
        console.log(replaceString)
        expression = expression.replace(toBeReplaced, replaceString)
        indexOfFraction = expression.indexOf("/")
    }
    while (expression.indexOf("*") != -1) {
        expression = expression.replace("*", "")
    }
    
    return expression;
}

var main = function() {
    console.log("Ready");
    
    $("#info").click(function() {
        $("#matrix").empty();
        rows = $("#rows").val();
        columns = $("#columns").val();
        $("#variables").css("display", "block")
        
        if (rows > 5) {
            for (i = 0; i < rows; i++) {
                for (j = 0; j < columns; j++) {

                    $("#matrix").append("<input type='text' class='matrix-input' style='width: 90px' id='c" + i + j + "'>")
                }
                $("#matrix").append('<br><br><br><br>')
            }
        } else {
            for (i = 0; i < rows; i++) {
                for (j = 0; j < columns; j++) {

                    $("#matrix").append("<input type='text' class='matrix-input' id='c" + i + j + "'>")
                }
                $("#matrix").append('<br><br><br><br>')
            }
        }
        
        $("#matrix").append('<span id="matrixsubmit" class="yolo">Submit Matrix</span>')   

        main();
    })
    
    $("#matrixsubmit").click(function() {
        $(".result").empty();
        var variablesStr = $("#variables-input").val();
        if (variablesStr != "") {
            console.log(variablesStr)
            variablesStr = matlabFriendly(variablesStr);

            variablesArr = variablesStr.split(",")
            var MATLABVarCommand = "syms "
            $.each(variablesArr, function(index, value) {
                MATLABVarCommand += value
                MATLABVarCommand += " "
            })
            $(".result").append("<h1>MATLAB Variable Initializer</h1>")
            $(".result").append("<p>" + MATLABVarCommand + "</p>")
        }
        
        var matrix = []
        for (i = 0; i < rows; i++) {
            matrix.push([]);
            for (j = 0; j < columns; j++) {
                var id = '#c' + i + j
                console.log(id)
                if ($(id).val() == "") {
                    matrix[i][j] = 0
                } else {
                    matrix[i][j] = $(id).val().replace(/\s+/g, '');;
                }
            }
        }
        console.log(matrix);
        
        
        
        
        // MATLAB
        matlabMatrix = matrix
        commandm = "m = ["
        $.each(matlabMatrix, function(index, value) {
            
            $.each(value, function(index2, value2) {
                commandm += value2 + " "
            })
            commandm += "; "
        });
        commandm += "]"
        commandm = matlabFriendly(commandm)
        console.log(commandm)
        
        $(".result").append("<h1>MATLAB Matrix Creation Command</h1>")
        $(".result").append("<p>" + commandm + "</p>")
        
        
        // latex
        latexMatrix = matrix
        commandl = "\\bea\n\\left[\\begin{array}{"
        for (j = 0; j < columns; j++) {
            commandl += "c"
        }
        commandl += "}<br>"
        $.each(latexMatrix, function(index, value) {
            
            $.each(value, function(index2, value2) {
                commandl += convertAlgebraToLATEX(value2) + " & "
            })
            commandl = commandl.slice(0, -2);
            commandl += "\\\\" + "<br><br>"
        });
        
        commandl += "\\end{array}\\right]<br>\\eea"
        $(".result").append("<h1>LATEX Matrix Creation Command</h1>")
        $(".result").append("<p>" + commandl + "</p>")
        
        $(document).scrollTop($(document).height());

    })
        
        
        
}

$(document).ready(main);
\end{lstlisting}

\lstset{language=Python}
\subsection{A Python Program for Verifying PPT States}
main.py
\begin{lstlisting}[caption=main.py]
#!/usr/bin/env python

import pandas as pd
from scipy import linalg as LA
import numpy as np
import os
dir_path = os.path.dirname(os.path.realpath(__file__))

phi1 = pd.read_excel(f"{dir_path}/Matrix.xlsx", sheet_name="Phi1", header=None).values
phi2 = pd.read_excel(f"{dir_path}/Matrix.xlsx", sheet_name="Phi2", header=None).values
phi1pt = pd.read_excel(f"{dir_path}/Matrix.xlsx", sheet_name="Phi1_T", header=None).values
phi2pt = pd.read_excel(f"{dir_path}/Matrix.xlsx", sheet_name="Phi2_T", header=None).values

for i in range(0, 100):
    print()


phi1e_vals = sorted(np.round_(LA.eig(phi1)[0], 3))
phi1Te_vals = sorted(np.round_(LA.eig(phi1pt)[0], 3))

phi2e_vals = sorted(np.round_(LA.eig(phi2)[0], 3))
phi2Te_vals = sorted(np.round_(LA.eig(phi2pt)[0], 3))

phi1eigpos = True
for i in phi1e_vals:
    if i < 0:
        phi1eigpos = False
        break

phi2eigpos = True
for i in phi2e_vals:
    if i < 0:
        phi2eigpos = False
        break

phi1Teigpos = True
for i in phi1Te_vals:
    if i < 0:
        phi1Teigpos = False
        break

phi2Teigpos = True
for i in phi2Te_vals:
    if i < 0:
        phi2Teigpos = False
        break

print(f"{phi1eigpos} \nEigenvalues of phi1 are\n\n", phi1e_vals)
print(f"{phi1Teigpos} \nEigenvalues of phi1 partial transpose are\n\n", phi1Te_vals)
print(f"{phi2eigpos} \nEigenvalues of phi2 are\n\n", phi2e_vals)
print(f"{phi2Teigpos} \nEigenvalues of phi2 partial transpose are\n\n", phi2Te_vals)
print(f"Overall: {phi2eigpos == True and phi1eigpos == True and phi1Teigpos == True and phi2Teigpos == True}")



\end{lstlisting}
\subsection{A Python Program for Calculating Choi Matrices, Linear Maps, and Compositions}

\begin{lstlisting}[caption=main.py]
import numpy as np
import math
import pandas as pd
import ast
import os
import time

dimension = 4 # Dimension of the quantum system
input = [[0] * dimension for x in range(dimension)]
variables = []
for i in range(0, dimension):
    for j in range(0, dimension):
        input[i][j] = "x_" + str(i + 1) + str(j + 1)


# ============== ENTRY CLASS =============
class Entry:

    def  __init__(self, value=None, coefficientMatrix=None, inputMatrix=input):
    	# Initializer
        self.value = value
        self.coefficientMatrix = coefficientMatrix
        self.inputMatrix = inputMatrix
        if coefficientMatrix is not None:
            self.dimension = len(coefficientMatrix[0])

    def getC(self): # Get Coefficients
        return self.coefficientMatrix

    def getV(self): # Get Value (if it exists)
        if self.value is None:
            return 0
        return self.value

    def getFC(self): # Returns String or raise error
        if self.coefficientMatrix is not None and self.dimension is not None:
            expression = ""
            for i in range(0, self.dimension):
                for j in range(0, self.dimension):
                    if str(self.coefficientMatrix[i][j]) != "0":
                        if str(self.coefficientMatrix[i][j]) == "1":
                            expression += str(self.inputMatrix[i][j]) + "+"
                        else:
                            expression += (str(self.coefficientMatrix[i][j]) + "*" + str(self.inputMatrix[i][j]) + "+")

            return expression[:-1]
        else:
            raise ArithmeticError

    def computeEntry(self, x): # Returns float or raise error
        if self.coefficientMatrix is not None and self.dimension is not None:
            sum = 0
            for i in range(0, self.dimension):
                for j in range(0, self.dimension):
                    if self.coefficientMatrix[i][j] != 0:
                        sum += self.coefficientMatrix[i][j] * x[i][j]

            return sum
        else:
            raise ArithmeticError

    def rawRepresentation(self, x=None):
        if self.value is not None:
            return self.getV()
        elif self.coefficientMatrix is not None and self.dimension is not None:
            if x is None:
                return self.getFC()
            else:
                return self.computeEntry(x)
        else:
            return 0

    def __str__(self):
        return str(self.rawRepresentation())

# =========== MATRIX CLASS =============

class Matrix:
    def __init__(self, dimension):
    	# Initializer
        self.dimension = dimension
        self.matrix = np.array([[Entry()] * dimension for x in range(dimension)])

    def setMatrixN(self, m): #input: 2D array of numbers
        self.dimension = len(m[0])
        d = self.dimension
        for i in range(0, d):
            for j in range(0, d):
                self.setEntry(i, j, v=m[i][j])

    def setMatrixE(self, e): #input: 2D array of Entries
        self.dimension = len(e[0])
        d = self.dimension
        for i in range(0, d):
            for j in range(0, d):
                self.matrix[i][j] = e[i][j]



    def setEntry(self, i, j, v=None, c=None, inputM=input):
        self.matrix[i][j] = Entry(value=v, coefficientMatrix=c, inputMatrix=inputM)

    def printM(self):
        for i in range(0, dimension):
            print(f"Row {i+1}:")
            for j in range(0, dimension):
                print(self.matrix[i][j].rawRepresentation())

    def returnMInArray(self):
        returnM = [[0] * dimension for x in range(dimension)]
        for i in range(0, dimension):
            for j in range(0, dimension):
                returnM[i][j] = self.matrix[i][j].rawRepresentation()
                if returnM[i][j] == '':
                    returnM[i][j] = '0'

        return returnM






    def convertChoiToLinearMap(self, d1, d2): #Input: 2 integers, output: 1 matrix
        if d1 * d2 != self.dimension:
            raise ArithmeticError
        else:
            linearMap = Matrix(d2)
            for i in range(0, d1):
                si = i * d1
                ei = si + d1
                for j in range(0, d2):
                    sj = j * d2
                    ej = sj + d2
                    subM = self.matrix[i::d2, j::d2]


                    # print(f"For {i}{j} the matrix is")
                    s = ""
                    for l in subM:
                        for k in l:
                            s += str(k.rawRepresentation()) + " "
                        s += "\n"

                    # print(s)



                    coefficientMForThatEntry = Entry(coefficientMatrix=subM)
                    # print("Dim of sumM is ", len(subM), len(subM[0]))
                    # print(len(coefficientMForThatEntry.coefficientMatrix), " ", len(coefficientMForThatEntry.coefficientMatrix))
                    if coefficientMForThatEntry.rawRepresentation() == "":
                        coefficientMForThatEntry = Entry(value=0)

                    # print(f"For {i}{j} the equation is {coefficientMForThatEntry}\n\n\n\n")
                    linearMap.matrix[i][j] = coefficientMForThatEntry

            return linearMap





# ============== MATRIX OPERATIONS =============

def compositeM(phi1, phi2): #Input two matrices, output one matrix. This will be phi2(phi1(x))
    dimension = len(phi2.returnMInArray())
    m1 = phi2.returnMInArray()
    m2 = phi1.returnMInArray()
    for i in range(0, dimension):
        for j in range(0, dimension):
            m2[i][j] = str(m2[i][j]).replace("x", "z")
    for i in range(0, dimension):
        # For each row in m1
        for j in range(0, dimension):
            # For each row in m2
            for k in range(0, dimension):
                # For each item in m2, find replace its respective element in m1
                m1[i] = [str(s).replace(input[j][k], "(" + m2[j][k] + ")") for s in m1[i]]

    # Convert m1 to a Matrix object
    compositeMatrix = Matrix(phi1.dimension)
    for i in range(phi1.dimension):
        for j in range(phi1.dimension):
            compositeMatrix.setEntry(i, j, v=m1[i][j], inputM=phi1)

    return compositeMatrix


def compositeA(phi1, phi2):
    dimension = len(phi1)
    phi2Str = str(phi2).replace("x", "z")
    for i in range(0, dimension):
        for j in range(0, dimension):
            phi2Str = phi2Str.replace(input[i][j].replace("x", "z"), "(" + phi1[i][j] + ")")

    phi2Array = ast.literal_eval(phi2Str)
    return phi2Array

def linearMapToChoiMatrix(d1, d2, matrix):
    minput = [[0] * d1 for x in range(d1)]
    for i in range(0, d1):
        for j in range(0, d1):
            minput[i][j] = "x_" + str(i + 1) + str(j + 1)

    blankMatrix = np.zeros((d1 * d2, d1 * d2))
    for i in range(0, d1):
        for j in range(0, d1):
            # Big level
            beginCoordinate = (i * d1, j * d1)

            # Small level
            matrixString = str(matrix)

            matrixString = matrixString.replace("x_" + str(i + 1) + str(j + 1), "1")
            for m in minput:
                for n in m:
                    matrixString = matrixString.replace(n, "0")
            matrixArray = ast.literal_eval(matrixString)
            for m in range(0, d2):
                for n in range(0, d2):
                    # print(matrixArray[m][n])
                    blankMatrix[beginCoordinate[0] + m][beginCoordinate[1] + n] = eval(str(matrixArray[m][n]))

    return blankMatrix



p1 = [
    ["3*x_11+x_22", "1/2*x_21"],
    ["2*x_22", "4*x_11"]
]

print("Example of Converting Linear Map to Choi Matrix")
print(linearMapToChoiMatrix(2, 2, p1))
time.sleep(2)


print("\n\n\n\nExample of Converting Choi Matrix to Linear Map")
m = Matrix(dimension ** 2)
choi =[[1/4, 1/4, 0, 0, 1/4, 1/4, 0, 0, 0, 0, 0, 0, 0, 0, 0, -1/2], [1/4, 1/4, 0, 0, 1/4, 1/4, 0, 0, 0, 0, 0, 0, 0, 0, 0, -1/2], [0, 0, 0, 0, 0, 0, 0, 0, 0, 0, 0, 0, 0, 0, 0, 0], [0, 0, 0, 1/2, 0, 0, 0, 1/2, 0, 0, 0, 0, 0, 0, 0, 0], [1/4, 1/4, 0, 0, 1/4, 1/4, 0, 0, 0, 0, 0, 0, 0, 0, 0, -1/2], [1/4, 1/4, 0, 0, 1/4, 1/4, 0, 0, 0, 0, 0, 0, 0, 0, 0, -1/2], [0, 0, 0, 0, 0, 0, 0, 0, 0, 0, 0, 0, 0, 0, 0, 0], [0, 0, 0, 1/2, 0, 0, 0, 1/2, 0, 0, 0, 0, 0, 0, 0, 0], [0, 0, 0, 0, 0, 0, 0, 0, 0, 0, 0, 0, 0, 0, 0, 0], [0, 0, 0, 0, 0, 0, 0, 0, 0, 0, 0, 0, 0, 0, 0, 0], [0, 0, 0, 0, 0, 0, 0, 0, 0, 0, 1/4, 0, 0, 0, 0, 0], [0, 0, 0, 0, 0, 0, 0, 0, 0, 0, 0, 1/4, 0, 0, -1/2, 0], [0, 0, 0, 0, 0, 0, 0, 0, 0, 0, 0, 0, 1/2, 1/2, 0, 0], [0, 0, 0, 0, 0, 0, 0, 0, 0, 0, 0, 0, 1/2, 1/2, 0, 0], [0, 0, 0, 0, 0, 0, 0, 0, 0, 0, 0, -1/2, 0, 0, 1, 0], [-1/2, -1/2, 0, 0, -1/2, -1/2, 0, 0, 0, 0, 0, 0, 0, 0, 0, 1]]
m.setMatrixN(choi)
lm = m.convertChoiToLinearMap(4, 4)
lm.printM()

time.sleep(2)

print("\n\n\n\nExample of Composing Linear Map")
s = compositeM(lm, lm)
s.printM()
\end{lstlisting}


\newpage

\section*{Acknowledgments}
The author of this thesis, Ryan Jin, developed a particular interest for quantum physics since he was in seventh grade and began reading books and watching videos online. He is also highly passionate about computer science and later discovered quantum computation to be at the intersection of these two fields he enjoys. He then started to study quantum computation and information theory, and with the assistance of his instructor, learned the mathematical foundation of linear algebra and knowledge in quantum information. Being somewhat ambitious, Ryan Jin decided to take on an open problem in quantum information theory and discovered the PPT Squared Conjecture on a curated list of open quantum problems.

This article is written with sincere gratitude, especially to my instructor and supervisor Dr Yang, who provided valuable guidance in this research. He not only taught me the fundamentals of linear algebra and quantum information theory, but also walked me through the stages of writing this thesis. I have always wanted to explore quantum computation and information. He showed me a path.

Secondly, I would like to express immense gratitude to the organizers and sponsors of this competition, without whom I would not have the opportunity to participate in such a prestigious and phenomenal event.

Last but not least, I would like to thank my parents, who offered inexpressible support for my endeavor.

\bibliographystyle{unsrt}

\bibliography{PPTHighDim}

\begin{thebibliography}{10}

\bibitem{report2012}
M.~Ruskai, M.~Junge, D.~Kribs, P.~Hayden, and A.~Winter.
\newblock Operator structures in quantum information theory.
\newblock {\em Final report of Banff International Research Station workshop:
  Operator structures in Quantum Information Theory.}, 2012.

\bibitem{choi1982}
M.-D. Choi.
\newblock Positive linear maps.
\newblock {\em Operator Algebras and Applications (Kingston, 1980), Proc.
  Sympos. Pure Math. Amer. Math. Soc.}, 38(2):583--590, 2012.

\bibitem{sevag2010}
S.~Gharibian.
\newblock Strong np-hardness of the quantum separability problem.
\newblock {\em Quantum Information and Computation}, 10(3):343--360, 2010.

\bibitem{peres1996}
A.~Peres.
\newblock Separability criterion for density matrices.
\newblock {\em Phys. Rev. Lett.}, 77:1413, 1996.

\bibitem{tang1986}
W.-S. Tang.
\newblock On positive linear maps between matrix algebras.
\newblock {\em Linear algebra and its applications}, 79:33--44, 1986.

\bibitem{kye2013}
S.-H. Kye.
\newblock Facial structures for various notions of positivity and applications
  to the theory of entanglement.
\newblock {\em Reviews in Mathematical Physics}, 25(02):1330002, 2013.

\bibitem{choi1972}
M.-D. Choi.
\newblock Positive linear maps on {$C^*$}-algebras.
\newblock {\em Canad. Math. J.}, 24:520--529, 1972.

\bibitem{chw2019}
Matthias Christandl, Alexander~M\"uller Hermes, and Michael~M. Wolf.
\newblock When do composed maps become entanglement breaking?
\newblock {\em Annales Henri Poincar\'e}, 20.

\bibitem{cyt2019}
L.~Chen, Y.~Yang, and Waishing Tang.
\newblock Positive-partial-transpose square conjecture for $n=3$.
\newblock {\em Physical Review A}, 99:012337, 2019.

\bibitem{lg2015}
L.~Lami and V.~Giovannetti.
\newblock Entanglement-breaking indices.
\newblock {\em Journal of Mathematical Physics.}, 56(9):092201, 2015.

\bibitem{lg2016}
L.~Lami and V.~Giovannetti.
\newblock Entanglement-saving channels.
\newblock {\em Journal of Mathematical Physics.}, 57:032201, 2016.

\bibitem{kmp2017}
Matthew Kennedy, Nicolas~A. Manor, and Vern~I. Paulsen.
\newblock Composition of {PPT} maps.
\newblock {\em Quantum Inf. Comput. 18}, 18(5-6):472--480, 2018.

\bibitem{rjp2018}
Mizanur Rahaman, Sam Jaques, and Vern~I. Paulsen.
\newblock Eventually entanglement breaking maps.
\newblock {\em Journal of Mathematical Physics}, 59(6):062201, 2017.

\bibitem{hrf2019}
E.-P. Hanson, C.~Rouz\'e, and D.-S. Franca.
\newblock On entanglement breaking times for quantum markovian evolutions and
  the $ppt^2$ conjecture.
\newblock {\em arXiv:1902.08173.}, 2019.

\bibitem{hhhh2009}
R.~Horodecki, P.~Horodecki, M.~Horodecki, and K.~Horodecki.
\newblock Quantum entanglement.
\newblock {\em Review of Modern Physics.}, 81:865--931, 2009.

\bibitem{sbl2001}
Anna Sanpera, Dagmar Bru\ss{}, and Maciej Lewenstein.
\newblock Schmidt-number witnesses and bound entanglement.
\newblock {\em Phys. Rev. A}, 63:050301, Apr 2001.

\bibitem{bchhkls2002}
D.~Bru\ss, J.-I. Cirac, P.~Horodecki, F.~Hulpke, B.~Kraus, M.~Lewenstein, and
  A.~Sanpera.
\newblock Reflections upon separability and distillability.
\newblock {\em Journal of Modern Optics.}, 49(8), 2002.

\bibitem{stm2010book}
E.~St\o rmer.
\newblock {\em Positive Linear Maps of Operator Algebras}.
\newblock Springer Monographs in Mathematics, Berlin, 2013.

\bibitem{cyz2018}
Benoit Collins, Zhi Yin, and Ping Zhong.
\newblock The ppt squared conjecture holds generically for some classes of
  independent states.
\newblock {\em Journal of Physics A: Mathematical and Theoretical}, 51(42).

\bibitem{agkl2010}
R.~Augusiak, J.~Grabowski, M.~Ku\'s, and M.~Lewenstein.
\newblock Searching for extremal ppt entangled states.
\newblock {\em Optics Communications.}, 283(5):805--813, 2010.

\end{thebibliography}

\end{document}